%% file: main.tex
\documentclass[aps,prl,twocolumn,superscriptaddress,a4paper,english]{revtex4-2}
\usepackage{times}
\usepackage{color}
\usepackage[colorlinks=true,urlcolor=blue,citecolor=blue]{hyperref}
\usepackage{url}
\usepackage{breakurl}
\usepackage{soul}
\usepackage{graphicx}
\usepackage{amsmath}
\usepackage{subfigure}
\usepackage{xcolor}
\usepackage{amssymb}
\usepackage{latexsym}
\usepackage{hyperref}% add hypertext capabilities
\DeclareGraphicsExtensions{.png,.eps}
\usepackage{float}
\usepackage{makecell}
\usepackage{diagbox}
\usepackage{setspace}
\usepackage{appendix}
\usepackage{etoolbox}

\newcommand {\R}{\textcolor {black}}
\newcommand {\RR}{\textcolor {black}}

\usepackage{xcolor}
\usepackage[urlcolor=blue]{hyperref}      % links to citations
\hypersetup{
    colorlinks = true,                    % text and not border
    citecolor = {blue},
    linkcolor = {purple},
           }

\begin{document}
% \linenumbers

\title{Persistent Ballistic Entanglement Spreading with Optimal Control in Quantum Spin Chains}

\author{Ying Lu}
\affiliation{Center for Quantum Physics and Intelligent Sciences, Department of Physics, Capital Normal University, Beijing 10048, China}
\author{Pei Shi}
\affiliation{Center for Quantum Physics and Intelligent Sciences, Department of Physics, Capital Normal University, Beijing 10048, China}
\author{Xiao-Han Wang}
\affiliation{Center for Quantum Physics and Intelligent Sciences, Department of Physics, Capital Normal University, Beijing 10048, China}
\author{Jie Hu}%\email[Corresponding author. Email: ]{jie.hu@cnu.edu.cn}
\affiliation{Center for Quantum Physics and Intelligent Sciences, Department of Physics, Capital Normal University, Beijing 10048, China}
\author{Shi-Ju Ran}\email[Corresponding author. Email: ]{sjran@cnu.edu.cn}
\affiliation{Center for Quantum Physics and Intelligent Sciences, Department of Physics, Capital Normal University, Beijing 10048, China}
\date{\today}

\begin{abstract}
% Based on the Lieb-Robinson bound, we can deduce that entanglement growth cannot exceed a linear rate over time, thus imposing a speed limit on quantum information propagation. 

Entanglement propagation provides a key routine to understand quantum many-body dynamics in and out of equilibrium. \R{The entanglement entropy (EE) usually approaches to a sub-saturation known as the Page value $\tilde{S}_{P} =\tilde{S} - dS$ (with $\tilde{S}$ the maximum of EE and $dS$ the Page correction) in, e.g., the random unitary evolutions. The ballistic spreading of EE usually appears in the early time and will be deviated far before the Page value is reached.} In this work, we uncover that \R{the magnetic field that maximizes the EE robustly induces} persistent ballistic spreading of entanglement in quantum spin chains. \R{The linear growth of EE is demonstrated to persist till the maximal $\tilde{S}$ (along with a flat entanglement spectrum) is reached. The robustness of ballistic spreading and the enhancement of EE under such an optimal control are demonstrated, considering particularly perturbing the initial state by random pure states (RPS's). These are argued as the results from the endomorphism of the time evolution under such an entanglement-enhancing optimal control for the RPS's.}
\end{abstract}

\maketitle

Quantum entanglement is a fundamental concept to reveal the essence of quantum systems in contrast to classical ones~\cite{bell_aspect_2004, nielsen_chuang_2010}. The dynamics of quantum entanglement under unitary time evolution provides a key routine to investigating the exotic phenomena and properties of quantum many-body physics, such as quasiparticle excitations~\cite{calabrese_evolution_2005, bastianello_entanglement_2020}, information propagation~\cite{jurcevic_quasiparticle_2014, clark_quantum_2014, luitz_information_2017, zhu_entanglement_2020}, many-body localization~\cite{znidaric_many-body_2008, bardarson_unbounded_2012, serbyn_universal_2013, vosk_many-body_2013, huse_phenomenology_2014, PhysRevB.97.214202, PhysRevLett.122.040606, de_tomasi_algebraic_2019}, and causality~\cite{schneider_spreading_2021}. 

Among the novel phenomena in quantum many-body dynamics, the emergence of ballistic transport of entanglement \R{attacks many attentions}. It is mostly observed in the integrable models, indicating the presence of quasiparticle propagations~\cite{calabrese_evolution_2005, chiara_entanglement_2006, hartman_time_2013}. Exceptions have been found where ballistic spreading appears in the diffusive nonintegrable systems~\cite{PhysRevLett.111.127205_2013, PhysRevX.7.031016, bianchi_linear_2018}. In both cases, the bipartite entanglement entropy (EE) grows linearly with time according to the Kardar-Parisi-Zhang equation~\cite{PhysRevLett.56.889,PhysRevB.100.125139}. \RR{Other examples exhibiting this property include the random unitary dynamics~\cite{Zyczkowski_1994, RUD23review} and Floquet spin models~\cite{PhysRevB.95.094206}.}

\RR{However, previous works suggest that the ballistic spreading persists generally in a relatively short time, and the EE eventually converges to a sub-saturation known as the Page value~\cite{PhysRevLett.71.1291}}
\begin{eqnarray}
	\tilde{S}_\text{P} = \tilde{S} - \frac{N_{\text{A}}}{2\ln{2} N_{\text{B}}} - O(\frac{1}{2^N}),
	\label{eq-Page}
\end{eqnarray}
\RR{where $\tilde{S} = \log_{2} 2^{N_{\text{A}}}=N_{\text{A}}$ is the maximum of EE, $N$ is the total number of spins, $N_\text{A}$ and $N_\text{B}$ are the numbers of spins in the two subsystems, respectively (with $N_{\text{A}}+N_{\text{B}}=N$; we take $N_{\text{A}} = N_{\text{B}}$ below for simplicity), and $O(\frac{1}{2^N})$ represents the rest higher-order contributions. Eq.~(\ref{eq-Page}) can be deduced from the random matrix theory. The final state is similar to a random pure state (RPS)~\cite{PhysRevLett.71.1291, PhysRevLett.111.127205_2013}. It is thus of theoretical and practical interests to seek for the dynamical processes where the EE exceeds the Page value with persistent ballistic spreading behavior during the evolution.}

\begin{figure}[tbp]
	\centering
	\includegraphics[angle=0,width=0.95\linewidth]{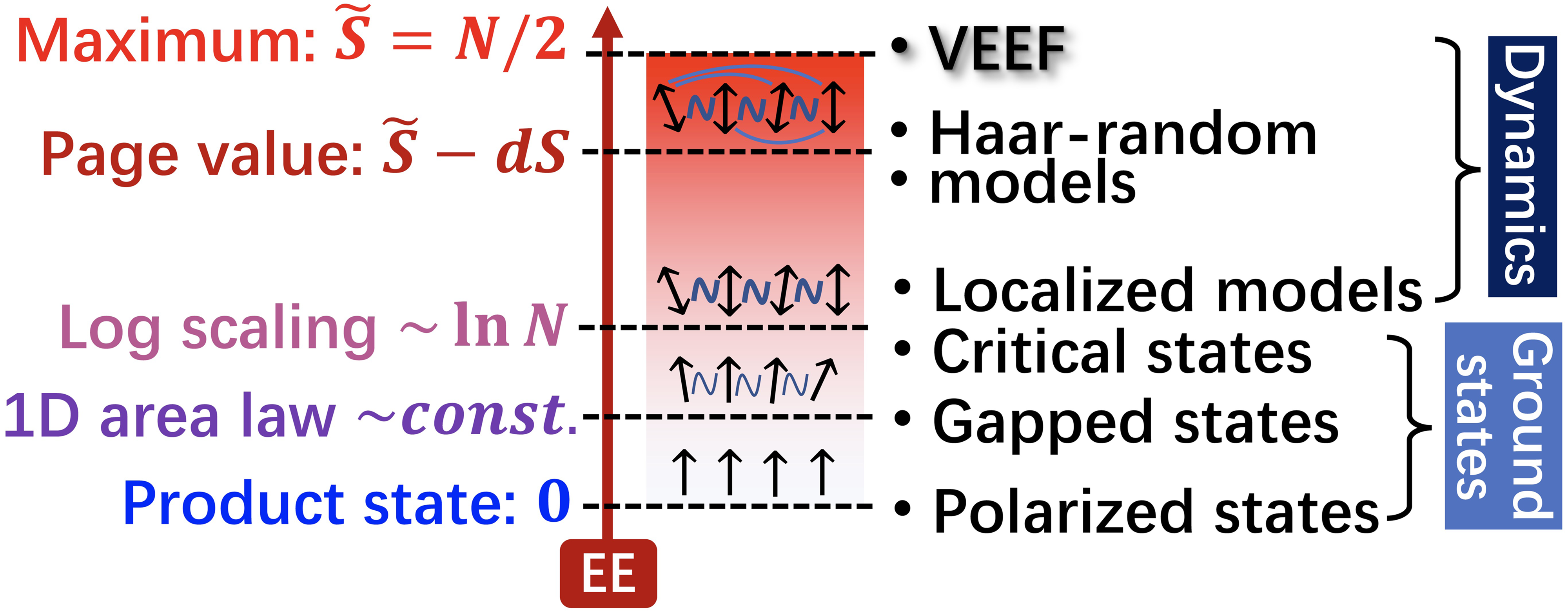}
	\caption{\R{(Color online) The illustration of different EE's (with equal bipartition) reachable by the 1D quantum systems with some typical examples listed in the right-hand side. The wavy lines among the arrows indicate the strength of entanglement. }}
	\label{fig-illu}
\end{figure}

In this work, we \R{show that quantum control by magnetic field can robustly reach the maximal $\tilde{S}$. Different EE's reachable by some typical 1D quantum systems are illustrated in Fig.~\ref{fig-illu}. The magnetic field is determined variationally by maximizing the EE in the center of the final state, which is dubbed as ``variational entanglement-enhancing” field (VEEF). Different from the previous works~\cite{hai_controlling_2010, zheng_controlling_2017},} ballistic spreading of entanglement is observed\R{, where the EE $S(t)$ at the time $t$ obeys
\begin{eqnarray}
	S(t) \simeq vt,
	\label{eq-linear}
\end{eqnarray}
with $v$ the velocity. The linear growth with VEEF will persist till the EE reaches $\tilde{S}=N/2$, when the evolution time $T$ satisfies
\begin{eqnarray}
	T \leq T_{\text{S}} \equiv \frac{N}{2v}.
	\label{eq-T}
\end{eqnarray}
\R{Note $T_{\text{S}}$ can be defined when the linear growth of EE can persist till $\tilde{S}$ is reached. In this case, the velocity defined as $v = dS(t)/dt$ is a constant for $t \leq T_S$.} \RR{For $T > T_{\text{S}}$, the EE still increases to the maximum $\tilde{S}$ at $t=T$. But during the evolution, the instantaneous velocity $v(t) \equiv dS(t)/dt$ is allowed to vary (in fact, one has $v(t) \leq v=\tilde{S} / T_{\text{S}}$, where $v$ can be considered as the maximal velocity that the system can reach). In the case of $T = T_{\text{S}}$ in comparison, the instantaneous velocity $v(t)$ has to be constantly $v$ so that the EE can reach $\tilde S = vT_{\text{S}}$ at $t=T$.} We also demonstrate the persistent ballistic spreading of EE on a localized model with VEEF. Generally, EE should spread sub-ballistically in the localized models~\cite{znidaric_many-body_2008,bardarson_unbounded_2012,swingle2013simple}.}

\R{The persistent ballistic spreading of EE is shown to be robust under random perturbations. A collapsing point is given by the EE curves from the initial states perturbed by different strengths of randomness. Such a behavior is explained and analyzed based on the random unitary evolution and the endomorphism of the VEEF evolution for the RPS's.}

\R{\textit{Variational entanglement-enhancing field.}--- Recent works revealed the inspiring prospective on applying machine learning (ML) methods in studying quantum dynamics and control~\cite{palittapongarnpim_learning_2017, yang_neural-network-designed_2018, bukov_reinforcement_2018, wu_learning_2019, zhang_when_2019, niu_universal_2019, schafer_differentiable_2020, zeng_quantum_2020, lu_preparation_2021, huang_machine-learning-assisted_2022}. Here, our aim is to enhance quantum entanglement by developing a ML-assisted quantum-control scheme. Enhancing entanglement~\cite{hai_controlling_2010, lubasch_dynamical_2011, zheng_controlling_2017, albarelli_locally_2018} is important} since entanglement is a fragile resource in noisy environments. 

We consider the time evolution with the Hamiltonian
\begin{eqnarray}
    \begin{aligned}
    \hat{H}(t) = \sum_{m,n} \hat{H}_{mn}
    + \sum_n \sum_{\alpha=x,z} h^{\alpha}_n(t) \R{\hat{\sigma}_n^{\alpha}}.
\label{eq-H}
    \end{aligned}
\end{eqnarray}
where $\hat{H}_{mn}$ represents the time-independent two-body interaction between the $m$-th and $n$-th sites, \R{$\hat{\sigma}_n^{\alpha}$ is the Pauli} operator ($\alpha=$x, y, z) on the $n$-th site, and $h^{\alpha}_n(t)$ denotes the time-dependent field. \R{The time evolution is a mapping}
\begin{eqnarray}
  \R{\mathcal{U}: |\psi_{0}\rangle \to |\psi(t)\rangle = e^{-i\int_{\tau=0}^{t}\hat{H}(\tau)d\tau} |\psi_{0}\rangle,}
\end{eqnarray}
\R{with $|\psi_{0}\rangle$ the initial state. Here, we consider $h^{\alpha}_n(t)$ as the variational parameters of $\mathcal{U}$, and optimize them by maximizing the EE $S$ of the final state $|\psi(T)\rangle$ (with $T$ the total evolution time). We dub the field satisfying the maximization condition $\partial S(T)/ \partial h^{\alpha}_n(t) = 0$ as variational entanglement-enhancing field (VEEF). In other words, the ``VEEF dynamics'' shows the properties of $\mathcal{U}$ by imposing $\partial S(T)/ \partial h^{\alpha}_n(t) = 0$.}

\R{To obtain VEEF, we adopt the} automatic differentiation technique that originated from the field of ML~\cite{JMLR:v18:17-468, lu_preparation_2021}. The time evolution is simulated by means of Trotter-Suzuki decomposition~\cite{trotter1959product, suzuki1976generalized}. \R{In the simulations, we take the \RR{Planck} constant $\hbar=1$ as the energy scale, and the time discretization for $h^{\alpha}_n(t)$ as $\tau =1/64\sim O(10^{-2})$ (which determines the highest frequency of $h^{\alpha}_n(t)$)}. With the bipartition to two halves denoted as $A$ and $B$, the EE \R{of $\vert\psi(t)\rangle$} satisfies
\begin{eqnarray}
	S(t) =-\mathrm{Tr}_A[\hat{\rho}(t)\log_2\hat{\rho}(t)],
	\label{eq-2}
\end{eqnarray}
with $\hat{\rho}(t)=\mathrm{Tr}_B|\psi(t)\rangle\langle\psi(t)|$ the reduced density matrix of $A$ by tracing over the degrees of freedom of $B$. The same results will be obtained by the reduced density matrix of $B$. \R{The maximal point is reached by} using the gradient descent method, $h^{\alpha}_n(t)$ as $h^{\alpha}_n(t) \leftarrow h^{\alpha}_n(t) + \eta \frac{\partial S(T)}{\partial h^{\alpha}_n(t)}$ with $S(T)$ the EE measured at the center of the final state. The gradients are obtained by the automatic differentiation technique and $\eta$ is the gradient step (or the learning rate in term of ML). To enhance the stability, we employ the fine-grained time optimization strategy~\cite{lu_preparation_2021}, and the ADAM optimizer~\cite{KB15Adam} that has been widely used in ML. 

Without losing generality, we take the initial state as a product state $|\psi_{0}\rangle = \prod_{\otimes n=1}^N |s_n\rangle$, where each spin $\vert s_n \rangle = \cos(\frac{\theta_n}{2})\vert0_n\rangle+e^{i\phi_n} \sin(\frac{\theta_n}{2})\vert 1_n\rangle$ points in a random direction on the Bloch sphere [$\theta_n\in[0,\pi$ and $\phi_n\in[0,2\pi)$]. The states $\vert 0_n\rangle$ and $\vert 1_n\rangle$ are the two eigenstates of \R{$\hat{\sigma}^{z}_{n}$}. Obviously, the initial state is not an eigenstate of the Hamiltonian. 

\begin{figure}[tbp]
	\centering
	\includegraphics[angle=0,width=0.95\linewidth]{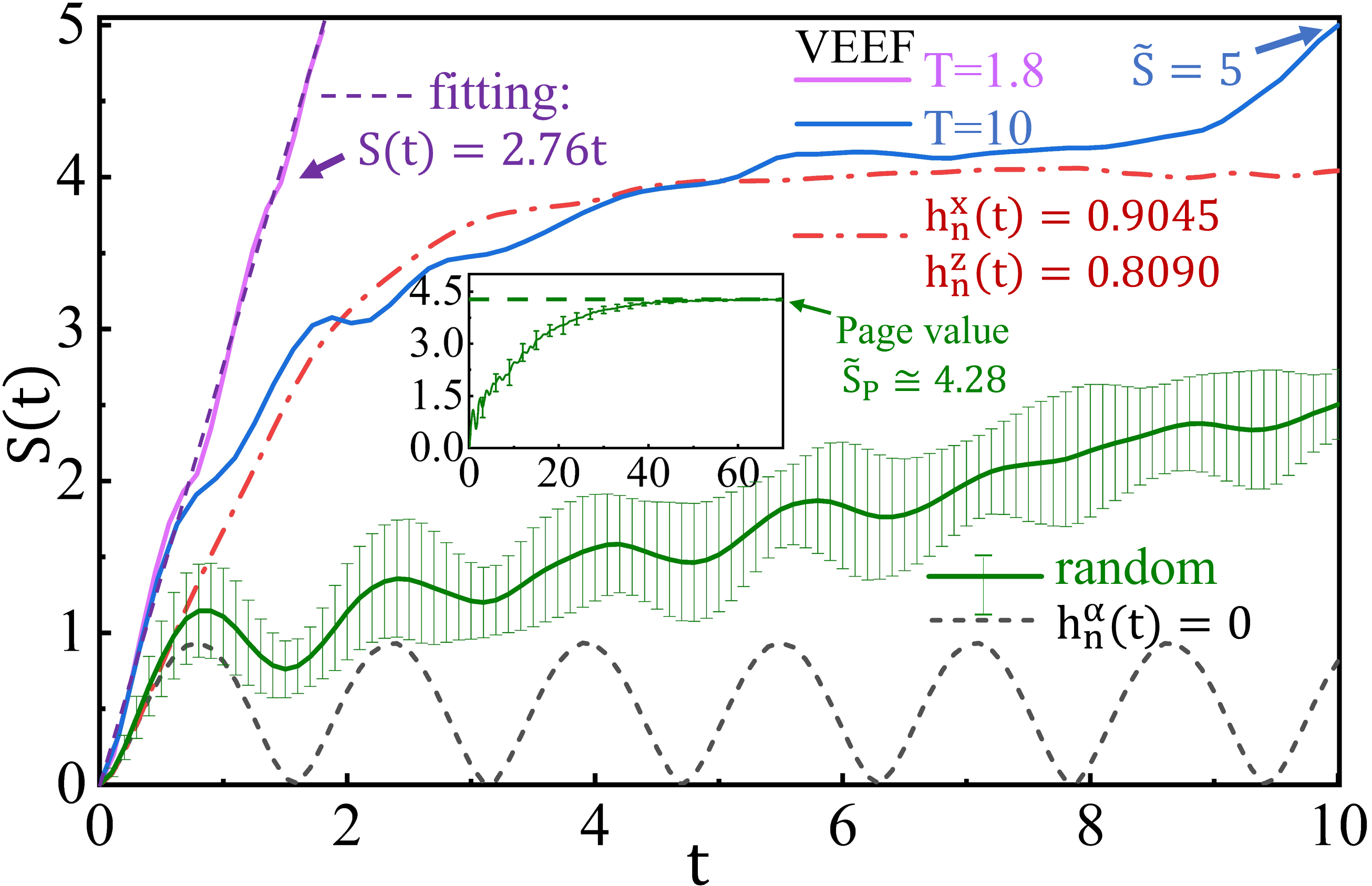}
	\caption{(Color online) The EE $S(t)$ against time $t$ with zero field $h^{\alpha}_n(t)=0$ (black dashed line), the constant field $h^x_n(t)=0.9045$ and $h^z_n(t)=0.8090$ where the system is nonintegrable~\cite{PhysRevLett.111.127205_2013} (red dash-dot line), random $h^{\alpha}_n(t)$ (green solid line), and VEEF ($T=1.8$ by purple solid line and $T=10$ by blue solid line). \R{We here consider the $N=10$ quantum Ising chain with periodic boundary condition.} The inset shows that EE converges to the Page value [Eq.~(\ref{eq-Page})] in the long-time limit~\cite{PhysRevLett.71.1291}. The results with the random $h^{\alpha}_n(t)$ are estimated by implementing ten independent simulations, where the variance is indicated by the error bars.}
	\label{fig-1}
\end{figure}

\textit{Persistent ballistic spreading of entanglement.---} We consider the one-dimensional (1D) quantum Ising model (QIM) with periodic boundary condition, whose Hamiltonian can be written as $ \hat{H}_{\text{QIM}}(t) =  \sum_{n=1}^{N-1}  \R{\hat{\sigma}_n^{z} \hat{\sigma}_{n+1}^{z} + \hat{\sigma}_1^{z} \hat{\sigma}_{N}^{z}} + \sum_{n=1}^N \sum_{\alpha=x,z} h^{\alpha}_n(t) \R{\hat{\sigma}_n^{\alpha}}$. The field is restricted to the spin x and spin z directions, which is a common scenario in theoretical and experimental investigations~\cite{doi:10.1126/science.1180085, PhysRevLett.110.084101}.

Figure.~\ref{fig-1} demonstrates the $S(t)$ [Eq. (\ref{eq-2})] with $N=10$ spins with the field taken in different ways. \R{The VEEF-driven spreading exhibits persistent linear growth until the maximal EE $\tilde{S}$ is reached (see the purple solid line in Fig.~\ref{fig-1} with \RR{$T=1.8 \simeq T_{\text{S}}$}). \R{This means $\partial S(T)/ \partial h^{\alpha}_n(t) = 0$ ($\sim O(10^{-8})$ from our numeric results) is satisfied for any $t\leq T_{S}$, consistent with the persistency of linear EE spreading with VEEF and the fact that 1D chains can at most exhibit linear growth of EE. Note $\tilde{S}$ corresponds to equal Schmidt coefficients ($\Lambda_{1} = \Lambda_{2} = \cdots $). The matrization of the state coefficients with $S=\tilde{S}$ gives a unitary matrix. \RR{Such a state can be regarded as the Choi state of a unitary operator~\cite{PhysRevA.101.052311}.} The measurement on one subsystem \RR{of the Choi state} will result in a unitary transformation on the collapsed state. In comparison, the Schmidt coefficients of the states whose EE's give the Page value deviate from being equal, particularly for moderately-large sizes~\cite{SMnote, PhysRevB.100.125115}.} We have $T_{\text{S}}=1.81$ by Eq.~(\ref{eq-T}) for the 1D QIM. The velocity $v=2.76$ with VEEF is much larger than $v=1.65$ obtained with the fixed field $h^x_n(t)=0.9045$, $h^z_n(t)=0.8090$ (where the system robustly becomes non-integral~\cite{Vnote, PhysRevLett.111.127205_2013, schachenmayer_entanglement_2013, PhysRevE.65.036208, PhysRevE.91.062128, PhysRevB.94.224202, Akila_2016, PhysRevB.95.094206}). The linear growth of $S(t)$ is only observed at the early time (for about $t < 3$) with such a fixed field.}

For $h^{\alpha}_n(t)=0$, all terms in the Hamiltonian commute with each other and $S(t)$ oscillates with $t$ far below the Page value~\cite{PhysRevB.95.094206}. With random $h^{\alpha}_n(t)$, $S(t)$ tends to increase over time, eventually approaching to the Page value in the long-time limit~\cite{PhysRevLett.71.1291} (about $t>60$ shown in the inset), behaving on average like a random pure state~\cite{PhysRevLett.111.127205_2013}. \R{In short, the previous means of implementing field reach the Page value of EE and induce a linear growth at the early time with a lower velocity.}

Extra degrees of freedom \R{emerges for the paths to a state with a maximal EE $\tilde{S}$ for $T>T_{\text{S}}$}. A consequence is that the linear growth is deviated before $\tilde{S}$ is achieved (see the blue solid line in Fig.~\ref{fig-1} with $T=10 \gg T_{\text{S}}$), \R{meaning $\partial S(T)/ \partial h^{\alpha}_n(t) = 0$ may not be satisfied in the duration $t < T$. But $\tilde{S}$ is robustly reached for the final state.}

\begin{figure}[tbp]
	\centering
	\includegraphics[angle=0,width=0.95\linewidth]{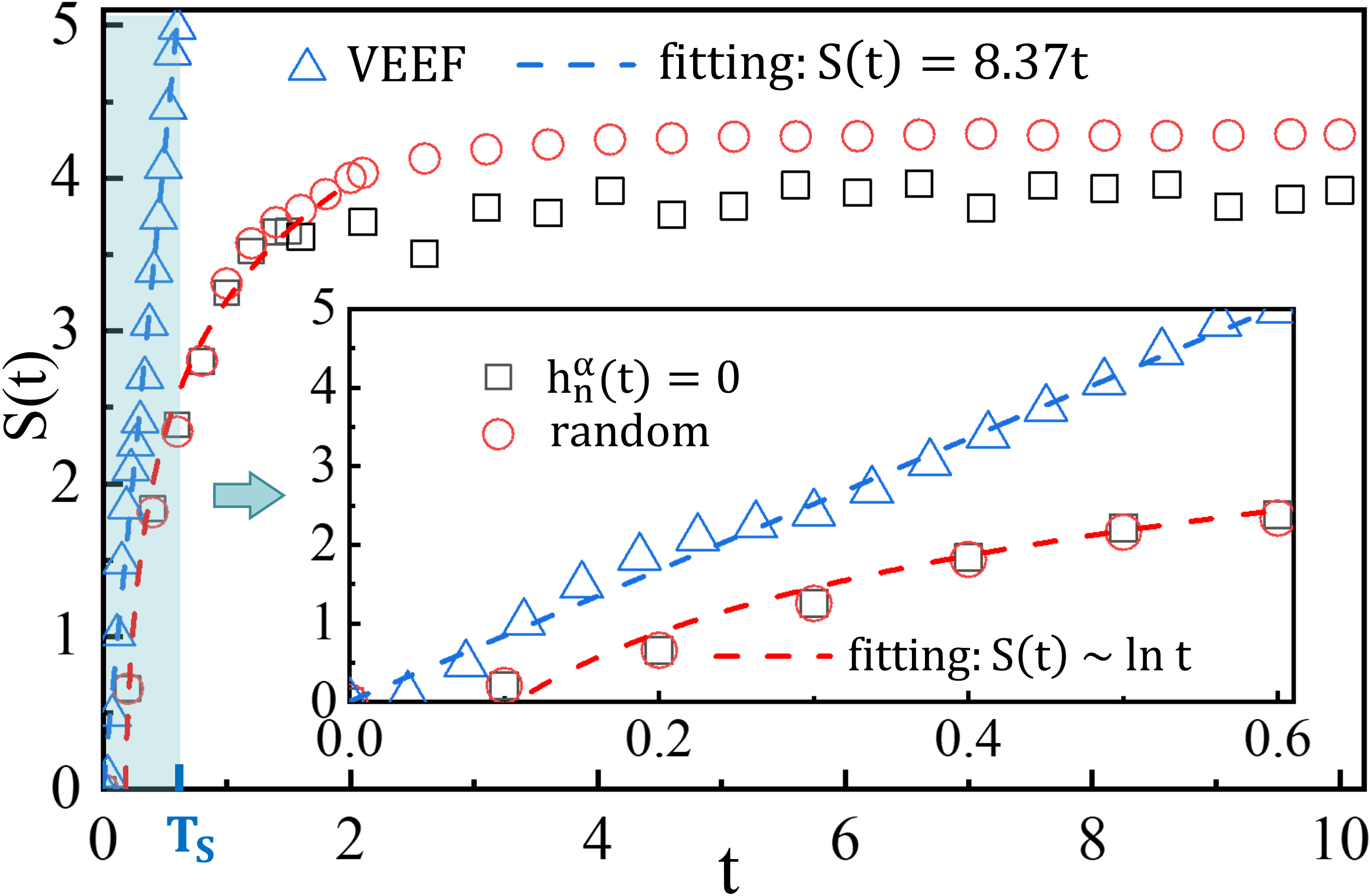}
	\caption{\R{(Color online) The EE $S(t)$ for the $N=10$ XXZ model obeys the logarithmic (diffusive) spreading with zero or random field~\cite{bardarson_unbounded_2012, znidaric_many-body_2008}, but the linear (ballistic) spreading with VEEF. The inset shows the zoom-in of the area for $t \leq T_{S}$.}}
	\label{fig-XXZ}
\end{figure}

\R{Fig.~\ref{fig-XXZ} demonstrates the ability of VEEF on changing the behaviors of EE spreading (that reflects the information spreading) from being diffusive to ballistic. The diffusive spreading of EE usually appears in the localized models~\cite{znidaric_many-body_2008,bardarson_unbounded_2012,swingle2013simple}. We take the XXZ chain with periodic boundary condition, where the local interaction satisfies $\hat{H}_{n,n+1} = \hat{\sigma}_n^{x} \hat{\sigma}_{n+1}^{x} + \hat{\sigma}_n^{y} \hat{\sigma}_{n+1}^{y} + \delta \hat{\sigma}_n^{z} \hat{\sigma}_{n+1}^{z}$ (with $\delta=3$ in Fig.~\ref{fig-XXZ}) and the VEEF is restricted in the x and z directions. With the zero or random field, diffusive spreading is expected, with $S(t) \sim \ln t$~\cite{znidaric_many-body_2008,bardarson_unbounded_2012}. With VEEF, linear growth $S(t) \sim vt$ with $v \simeq 8.37$ is obtained, which persists till $\tilde{S}$ is reached at $t=T_{s}$.}

\begin{figure}[tbp]
	\centering
	\includegraphics[angle=0,width=0.98\linewidth]{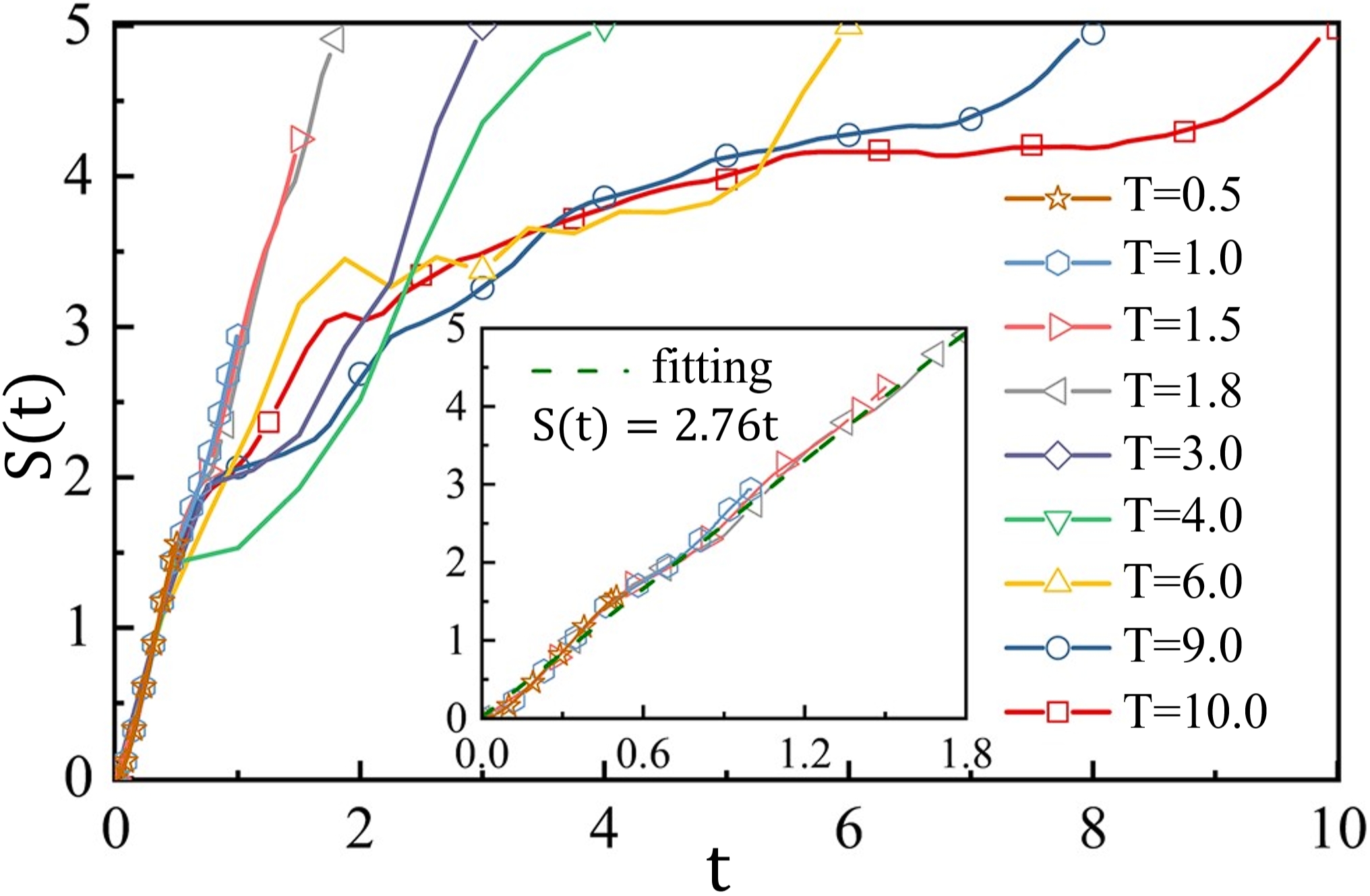}
	\caption{(Color online) The EE $S(t)$ \R{of the 1D QIM ($N=10$)} against the time $t$ with different total evolution time $T$. The inset shows the $S(t)$ for $T<T_{S}$, which satisfies the linear relation $S(t)=2.76t$.}
	\label{fig-2}
\end{figure}

\R{\textit{Robustness.---}} Figure.~\ref{fig-2} demonstrates the robustness of the ballistic EE spreading for different total evolution times $T$. The inset shows that the curves with different $T$'s ``perfectly'' collapse to the linear relation given by Eq.~(\ref{eq-linear}) \R{for the early times. In all cases, $\tilde{S}$ is reached for the final state.} \R{Note the states with the maximal EE are obviously not unique but form a subspace. The coefficients of a state therein is a unitary matrix from its Schmidt decomposition $\Psi = UV^{\dagger}/Z$ (with $Z$ denotes the normalization factor), where the Schmidt coefficients are equal.}

We conjecture that the whole subspace with maximal EE is in principle accessible by the VEEF for any $T \geq T_{\text{S}}$. To verify this conjecture, we take the target state as $|\psi(T')\rangle$ \R{with $T' \geq T_{\text{S}}$ and its EE $S=\tilde{S}$}, and try to evolve a random product state to $|\psi(T')\rangle$ in an evolution time \R{$T>T_{\text{S}}$}. The field is optimized by minimizing the infidelity $F_{\text{in}}(T; T') = 1 - | \langle \psi(T')| e^{-i\int_{\tau=0}^{t}\hat{H}(\tau)d\tau} |\psi(0) \rangle |$~\cite{lu_preparation_2021}. Taking $T=4$, $T'=6$ ($T<T'$) and $T=6$, $T'=4$ ($T>T'$) as two examples, vanishing infidelity with $F_{\text{in}}(T; T') \sim O(10^{-4})$ is obtained, which supports our conjecture. % With the VEEF that maximizes $S(T)$, we cannot specify which state in these subarea will be obtained, since the states therein have equal $S$ and thus the gradients of $S$ against $h^{\alpha}_{n}(t)$ should be zero.

\begin{figure}[tbp]
	\centering
	\includegraphics[angle=0,width=0.95\linewidth]{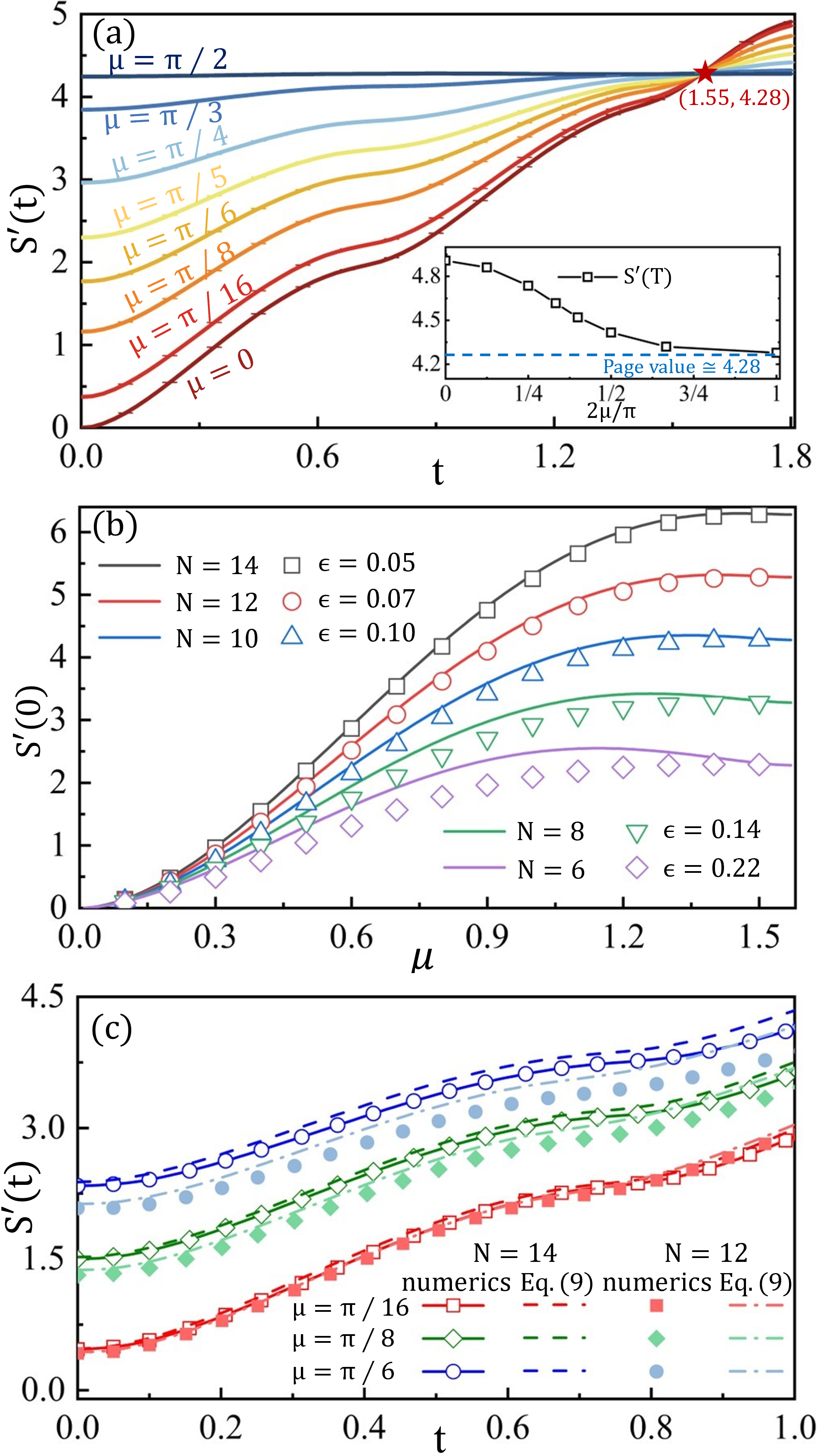}
	\caption{\R{(Color online) (a) The EE $S'(t)$ of 1D QIM against time $t$ for the initial states $\vert\psi'_{0}\rangle$ with with different $\mu$ controlling the strength of random perturbation [Eq.~(\ref{eq-psi1})]. $\vert\psi_{0}\rangle$ is the unperturbed initial state for obtaining the VEEF, and $\vert\psi_{r}\rangle$ is a RPS. The results are the average of $10$ independent simulations, and the error bars ($\sim O(10^{-3})$) are given by the variances. The insets show the $S'(T)$ versus $\mu$. (b) The EE $S'(0)$ of the perturbed initial state verses $\mu$ for different system sizes $N$, obtained by numerical simulations (symbols) and by Eq.~(\ref{eq-EEtime}) (solid lines). The violation of the orthogonal conditions is characterized by $\epsilon$. (c) The $S'(t)$ obtained by numerical simulations (symbols) and by Eq.~(\ref{eq-EEtime}) (solid lines) for $N=12, 14$ and $\mu = \pi/6, \pi/8, \pi/16$.}}
	\label{fig-4}
\end{figure}

\R{From the optimization process, VEEF should depend on the choice of the initial state $|\psi_{0}\rangle$. Fig.~\ref{fig-4} illustrates the robustness against random perturbations on $|\psi_{0}\rangle$ by
\begin{eqnarray}
	\vert\psi'_{0}\rangle = \cos \mu \vert\psi_{0}\rangle + \sin \mu \vert\psi_{\text{r}}\rangle,
	\label{eq-psi1}
\end{eqnarray}	
with $\mu$ controlling the strength of perturbation and $\vert\psi_{\text{r}}\rangle$ a RPS whose coefficients are randomly generated with a normal distribution. For $\mu=0$, no perturbation is added and the EE $S'(t)$ satisfies Eq.~(\ref{eq-linear}). For $\mu=\pi/2$, the evolution starts from a RPS with $S'(t=0) \simeq \tilde{S}_{\text{P}}$, and $S'(t)$ is almost a constant. This shows the time evolution $\mathcal{U}$ with VEEF is ``endomorphic'' for the RPS subspace, i.e., $\mathcal{U}_{\text{VEEF}}(|\psi_{\text{r}}\rangle) \in \text{RPS}$ for $|\psi_{\text{r}}\rangle \in \text{RPS}$. The VEEF evolution (optimized with a specific initial state) resembles the random unitary evolution when taking a RPS as the initial state.}

\R{For an arbitrary $\mu$, the perturbed initial state is the weighted superposition between the original initial state and a RPS. The perturbation preserves the linear behavior of EE but with lower velocities. This is reasonable since adjusting the field (one-body operators) does not alter the structure of the underlying quantum circuit, and therefore essentially does not change the Haar dynamics for the evolution of the RPS part (recall the endomorphism of $\mathcal{U}$ with VEEF, which maps a RPS to RPS). The data collapse at the crossing point of the two curves $S'(t)=vt$ [Eq.~(\ref{eq-linear}) with no perturbation] and $S'(t) = \tilde{S}_{\text{P}}$ (Page value with RPS). Consequently, we have $S'(T) \to \tilde{S}_{\text{P}}$ as $\mu \to \frac{\pi}{2}$ [see the inset of Fig.~\ref{fig-4} (a)]. This implies we could exceed the Page value $S'(T) > \tilde{S}_{\text{P}}$ by implementing VEEF on an initial state $| \psi'_{0} \rangle$ with $|\langle \psi_{0} | \psi'_{0} \rangle | > 0$ when the orthogonal part is described by a RPS. In this case, the perturbation will enlarge the EE [$S'(t) > S(t)$ with $S(t)$ the EE without perturbation] for the time duration before the collapsing point.}

\R{To further analyze $S'(t)$, we assuming the following orthogonal conditions between the Schmidt basis states of $\vert\psi_{0}\rangle$ and $\vert\psi_{\text{r}}\rangle$ as 
\begin{equation}
	\begin{aligned}
		{\ }_{i}\langle \psi^{\text{A}}(t) \vert \psi_{\text{r}}^{\text{A}}(t)\rangle_{i'} \sim 0, \ 
		{\ }_{i}\langle \psi^{\text{B}}(t) \vert \psi_{\text{r}}^{\text{B}}(t)\rangle_{i'} \sim 0. 
		\label{eq-fai-orthogonal}
	\end{aligned}
\end{equation}
with $\left\{\vert \psi^{\text{A}}(t) \rangle_i, \vert \psi^{\text{B}}(t) \rangle_i \right\}$ and $\left\{ \vert \psi_{\text{r}}^{\text{A}}(t) \rangle_i, \vert \psi_{\text{r}}^{\text{B}}(t) \rangle_i \right\}$ the left/right Schmidt basis states of $\mathcal{U}_{\text{VEEF}}(\vert\psi_{0}\rangle)$ and $\mathcal{U}_{\text{VEEF}}(\vert\psi_{\text{r}}\rangle)$, respectively. Then, the EE approximately obeys
\begin{equation}
	S'(t) \simeq \tilde{S}_{\mu} + \tilde{S}_{\text{P}} \sin^{2} \mu + S(t) \cos^{2} \mu,
    \label{eq-EEtime}
\end{equation}
where $\tilde{S}_{\mu} = -\cos^{2} \mu \log_{2} \cos^{2} \mu - \sin^{2} \mu \log_{2} \sin^{2} \mu$ can be treated as an additional entropy from a two-level probabilistic distribution $p(\mu) = [\cos^{2} \mu, \sin^{2} \mu]$~\cite{SMnote}. Eq.~(\ref{eq-EEtime}) implies the linear growth of $S'(t)$ since $S(t)$ satisfies Eq.~(\ref{eq-linear}).}

\R{At $t=0$, the EE of the initial state can be well predicted by Eq.~(\ref{eq-EEtime}) particularly for $N=14$, since the orthogonal conditions between a RPS and a product state can be well satisfied with a moderately large $N$ [see Fig.~\ref{fig-4}(b); note $S(0)=0$]. The violation of Eq.~(\ref{eq-fai-orthogonal}) is characterized by $\epsilon =\frac{1}{2D} \left(\sum_{i=1}^{D} \left| \langle\psi^{\text{A}}(0) \vert\psi_{\text{r}}^{\text{A}}(0)\rangle_{i} \right| + \sum_{i=1}^{D}\left| \langle\psi^{\text{B}}(0) \vert\psi_{\text{r}}^{\text{B}}(0)\rangle_{i} \right|\right)$, which should in general vanish exponentially in the $N \to \infty$ limit. For $t>0$, the deviation between the VEEF results and Eq.~(\ref{eq-EEtime}) is small for small $\mu$ and $t$, and for large $N$ [Fig.~\ref{fig-4} (c)]. For large $\mu$ or $t$, we essentially require the corresponding orthogonality between a RPS and an entangled state, which is more difficult to achieve and thus requires larger $N$~\cite{SMnote}. We expect Eq.~(\ref{eq-EEtime}) to be held for $N \to \infty$.}

\textit{Summary.---} \R{We have uncovered} the persistent ballistic spreading of entanglement under the variational entanglement-enhancing field that maximizes the entanglement entropy of the final state. \R{Be persistent, we mean that the linear growth of EE (with equal bipartition) holds till the maximal EE} is reached. This is in contrast to the previous results, where the EE generally converges to the Page value in the long-time limit and the linear growth just appears at the early time of evolution. \R{The robustness of the EE ballistic spreading under VEEF is investigated. When the perturbation is described by a random pure state (RPS), the VEEF can enlarge the EE in a predictable manner and its persistent linear growth can be preserved. This is a result of the endomorphism of the VEEF time evolution for the RPS's.}

%\section*{Acknowledgment} 
\textit{Acknowledgment.} We thank Ding-Zu Wang and Peng-Fei Zhou for helpful discussions. This work was supported in part by NSFC (Grants No. 12004266 and No. 11834014), Beijing Natural Science Foundation (Grant No. 1232025), Tianjin Natural Science Foundation of China (Grant No. 20JCYBJC00500), and Academy for Multidisciplinary Studies, Capital Normal University.

\bibliographystyle{apsrev4-2}
% \bibliography{name}
\input{main.bbl}

\clearpage
\onecolumngrid
\appendix

\section{SUPPLEMENTAL MATERIAL OF ``Persistent Ballistic Entanglement Spreading with Optimal Control in Quantum Spin Chains"}
\input{name_new}

\end{document}

%% file: main.bbl
%apsrev4-2.bst 2019-01-14 (MD) hand-edited version of apsrev4-1.bst
%Control: key (0)
%Control: author (72) initials jnrlst
%Control: editor formatted (1) identically to author
%Control: production of article title (-1) disabled
%Control: page (0) single
%Control: year (1) truncated
%Control: production of eprint (0) enabled
%

%% file: name_new.tex
% \section{Quantum circuit encapsulation}
In this supplemental material, we provide more analyses on the ballistic spreading of entanglement entropy (EE) induced by the ``variational entanglement-enhancing” field (VEEF). We also provide more information on the robustness of such a ballistic EE spreading. Finally, the data of VEEF on the transverse Ising model is provided as an example.

\section{I. Basic setup}

We consider the time-dependent Hamiltonian to have the form as
\begin{eqnarray}
	\begin{aligned}
		\hat{H}(t) = \sum_{m,n} \hat{H}_{mn}
		+ \sum_n \sum_{\alpha=x,z} h^{\alpha}_n(t) \hat{\sigma}_n^{\alpha}.
		\label{eq-H}
	\end{aligned}
\end{eqnarray}
where $\hat{H}_{mn}$ represents the time-independent two-body interaction between the $m$-th and $n$-th sites, $\hat{\sigma}_n^{\alpha}$ is the Pauli operator ($\alpha=$ x, y, z) on the $n$-th site, and the field $h^{\alpha}_n(t)$ is the function of time that is to be determined variationally. 

Our goal is to optimize $h^{\alpha}_n(t)$ for $0 \leq t \leq T$ by maximizing the entanglement entropy (EE) $S$ of the final state $\vert\psi(T)\rangle =e^{-i\int_{\tau=0}^{T}\hat{H}(\tau)d\tau}\vert\psi_0\rangle$, with $\vert\psi_0\rangle$ the initial state. Without generality, we consider the equal bipartition. The field at the maximal point satisfies $\delta S(T)/ \delta h^{\alpha}_n(t) = 0$, and is dubbed as ``variational entanglement-enhancing field" (VEEF). By discretizing the time using Trotter-Suzuku decomposition and not requiring $h^{\alpha}_n(t)$ to be continuous against $t$ (which is feasible in experiments, akin to the bang-bang protocols), the functional differentiations can be replaced by the partial differentiations. The partial differentiations can be efficiently computed by the automatic differentiation technique.

\R{The essential differences between VEEF and the variational methods such as time-dependent variational principle (TDVP)~\cite{Dirac_1930, PhysRevLett.93.207204} can be seen from the target function and the variational parameters. Taking the TDVP based on matrix product state (MPS) as an example, its target is essentially to minimize the error in simulating the dynamical process given a Hamiltonian. The dynamical process (say the evolving state) is parameterized as MPS, where the tensors forming the MPS will be optimized during the simulation. For VEEF, the main aim is to maximize a target property (such as the EE of the final state) by variationally optimizing some parameters of the dynamical system (such as the magnetic field in the Hamiltonian). In other words, the variational parameters to be adjusted in VEEF are not to give the dynamics but the Hamiltonian. Thus, we believe VEEF would offer a unique perspective to understand quantum many-body dynamics.}

The results given in the main text show that the VEEF induces persistent ballistic spreading of entanglement in quantum spin chains with a linear growth in EE until it reaches the maximum $\tilde{S} = - \log_{2} 2^{-\frac{N}{2}}=\frac{N}{2}$ with $N$ the total number of spins. For $t \leq \frac{N}{2v}$, the EE satisfies $S(t) = v t$ with $v$ the velocity. These results are in sharp contrast with the behaviors without VEEF, where the EE generally approaches a sub-saturation known as the Page value $\tilde{S}_{P} =\tilde{S} - \frac{1}{2\ln{2}}$ in the long-time limit, and the entanglement growth deviates from being linear before the Page value is reached. 

\R{The complexity of the numerical computation for VEEF is high. Though handling the time evolution with a sparse Hamiltonian at every time window can be quite efficient, one key reason for the high complexity of VEEF computation is the cost on the field optimization procedure. Meanwhile, the high complexity more essentially concerns representing the time-dependent state $| \psi (t)\rangle$, which becomes very dense as its EE grows. Since $| \psi (t)\rangle$ obeys the volume law (instead of area law) of EE, the complexity of accurately representing it scales exponentially with the system size. Even we do not do any optimization, saving such a state in a computer is exponentially expensive. Note $| \psi (t)\rangle$ obeys the volume law particularly when the time $t$ approaches to the total evolution time $T$, because the VEEF maximizes the EE of the final state $| \psi (T)\rangle$. Therefore, the exponential complexity from the volume law seems inevitable in our simulation, as the EE of $| \psi (t)\rangle$ robustly reaches the maximum under VEEF.}

% \R{In conclusion, VEEF can be viewed as a variational method that distinguishes it from traditional approaches such as TDVP. While TDVP (time-dependent variational principle) is a quantum state solving method originally proposed by Dirac in discussions on the evolution of charge distributions~\cite{Dirac_1930}. It can be applied to various variational wave functions, such as Hartree-Fock, but the current trend favors the use of a specific form like Matrix Product States (MPS)~\cite{PhysRevLett.93.207204, PhysRevLett.107.070601}. The core idea of TDVP is to project the time evolution of the system onto the tangent space of the variational wave function, aiming to find the optimal description of the system's time evolution process.}

% \R{TDVP is widely used for determining a system's time evolution by minimizing the energy functional and excels in handling various types of Hamiltonians and quantum systems, it may face challenges in dealing with highly complex systems in terms of computational efficiency and scalability. In contrast, VEEF focuses on studying the impact of single-body operators in the Hamiltonian on quantum system dynamics, aiming to maximize specific physical properties through parameter optimization within the Hamiltonian. This approach offers a unique perspective and potential advantages for efficiently understanding and manipulating quantum system dynamics.}

\section{II. Entanglement entropy perturbed by random pure state}

\R{Though the high complexity of VEEF brings the restrictions on the accessible system size, we below provide some analytical results to reveal some properties in the large-size limit, where the accuracies of the analytical predictions increase with the system size (some numerical results are provided in Fig.~5 of the main text).}

The main text illustrates the robustness of the EE ballistic spreading against random perturbations on the initial product state $|\psi_{0}\rangle$. Specifically, we perturb it by adding a random pure states (PRS) $\vert\psi_{\text{r}}\rangle$ as
\begin{equation}
	\begin{aligned}
		\vert\psi'_{0}\rangle & = \cos\mu \vert\psi_{0}\rangle + \sin\mu \vert\psi_{\text{r}}\rangle,
		\label{eq-fai0}
	\end{aligned}
\end{equation}
with $\mu \in [0, \pi/2]$ controlling the strength of perturbation.

To analyze the $S(t)$ with the above perturbation, we assume some orthogonal conditions on the Schmidt basis states of $\vert\psi_{0}\rangle$ and $\vert\psi_{\text{r}}\rangle$. The Schmidt decompositions of these two states can be formally written as
\begin{equation}
	\begin{aligned}
		\vert\psi_{0}\rangle = \vert\psi_{0}^{\text{A}}\rangle\otimes\vert\psi_{0}^{\text{B}}\rangle,  \  \ 
		\vert\psi_{\text{r}}\rangle = \sum_{i=1}^{D}\tilde{\Lambda}_{i} \vert\psi_{\text{r}}^{\text{A}}\rangle_{i}\otimes\vert\psi_{\text{r}}^{\text{B}}\rangle_{i}.
		\label{eq-fai0-sd}
	\end{aligned}
\end{equation}
According to the properties of RPS, the Schmidt coefficients $\{\tilde{\Lambda}_{i}\}$ should give an EE of the Page value. Substituting above equation into Eq.~(\ref{eq-fai0}), we have
\begin{equation}
	\begin{aligned}
		\vert\psi'_{0}\rangle = \cos\mu \vert\psi_{0}^{\text{A}}\rangle\otimes\vert\psi_{0}^{\text{B}}\rangle + 
		\sin\mu\sum_{i=1}^{D} \tilde{\Lambda}_{i}\vert\psi_{\text{r}}^{\text{A}}\rangle_{i}\otimes\vert\psi_{\text{r}}^{\text{B}}\rangle_{i},
		\label{eq-fai0'}
	\end{aligned}
\end{equation}

Now, we assume the following orthogonal conditions
\begin{equation}
	\begin{aligned}
		\langle\psi_{0}^{\text{A}}\vert\psi_{\text{r}}^{\text{A}}\rangle_{i} \sim 0, \ \ 
		\langle\psi_{0}^{\text{B}}\vert\psi_{\text{r}}^{\text{B}}\rangle_{i} \sim 0,
		\label{eq-fai0-or}
	\end{aligned}
\end{equation}
These conditions should generally be satisfied in the large-size limit ($N \to \infty$)\R{, meaning the accuracy of the analytical prediction will increase with $N$}. Then the Schmidt bases of $\vert\psi'_{0}\rangle$ is the direct sum of those of $\vert\psi_{0}\rangle$ and $\vert\psi_{\text{r}}\rangle$, and the Schmidt coefficients of $\vert\psi'_{0}\rangle$ satisfy
\begin{equation}
	\begin{aligned}
		\boldsymbol{\Lambda}' = [\cos \mu, \ \tilde{\Lambda}_{1} \sin \mu, \ \tilde{\Lambda}_{2} \sin \mu, \cdots]
		\label{eq-fai0-es}
	\end{aligned}
\end{equation}
Therefore, the EE of $\vert\psi'_{0}\rangle$ satisfies
\begin{equation}
	\begin{aligned}
		S'(0)& = -\sum_{i=1}^{D+1} (\Lambda'_{i})^{2} \log_{2} (\Lambda'_{i})^{2} &\\
		& = -\sin^2\mu \log_{2}\sin^2\mu \sum_{i=1}^{D}(\tilde{\Lambda}_{i})^{2}  -\cos^2\mu \log_{2} \cos^2\mu -\sin^2\mu \sum_{i=1}^{D} (\tilde{\Lambda}_{i})^{2} \log_{2} (\tilde{\Lambda}_{i})^{2}.
		\label{eq-fai0-EE}
	\end{aligned}
\end{equation}
With $\sum_{i=1}^{D}(\tilde{\Lambda}_{i})^{2}=1$ and $-\sum_{i=1}^{D} (\tilde{\Lambda}_{i})^{2} \log_{2} (\tilde{\Lambda}_{i})^{2} = \tilde{S}_{P}$, we finally have
\begin{equation}
	\begin{aligned}
		S'(0) = \tilde{S}_{\mu} + \tilde{S}_{\text{P}} \sin^{2} \mu,
		\label{eq-fai0-EE'}
	\end{aligned}
\end{equation}
where we have an additional entropy
\begin{equation}
	\begin{aligned}
		\tilde{S}_{\mu} = -\sin^2\mu \log_{2} \sin^2\mu -\cos^2\mu \log_{2} \cos^2\mu.
		\label{eq-fai0-EEmu}
	\end{aligned}
\end{equation}
This can be treated as the entropy from a two-level probabilistic distribution $p(\mu) = [\cos^{2} \mu, \sin^{2} \mu]$. 

The EE of $\vert\psi'(t)\rangle  = e^{-i\int_{\tau=0}^{t}\hat{H}(\tau)d\tau} \vert\psi'_{0}\rangle$ in the time evolution is much more complicated to predict. Based on the linearity of the evolution, we have
\begin{equation}
	\vert\psi'(t)\rangle = \cos\mu \  e^{-i\int_{\tau=0}^{t}\hat{H}(\tau)d\tau} \vert\psi_{0}\rangle + \sin\mu \  e^{-i\int_{\tau=0}^{t}\hat{H}(\tau)d\tau} \vert\psi_{\text{r}}\rangle,
	\label{eq-fai}
\end{equation}

Denoting $\vert\psi(t)\rangle = e^{-i\int_{\tau=0}^{t}\hat{H}(\tau)d\tau} \vert\psi_{0}\rangle$ (the VEEF-evolution of the ``correct'' initial state) and $\vert\psi_{\text{r}}(t)\rangle = e^{-i\int_{\tau=0}^{t}\hat{H}(\tau)d\tau} \vert\psi_{\text{r}} \rangle$, their Schmidt decompositions can be formally written as
\begin{eqnarray}
	\vert\psi(t)\rangle = \sum_{i} \Lambda_{i}(t)\vert\psi^{\text{A}}(t)\rangle_{i}\otimes\vert\psi^{\text{B}}(t) \rangle_{i}, \ \ 
	\vert\psi_{\text{r}}(t)\rangle = \sum_{i} \tilde{\Lambda}_{i}(t) \vert\psi_{\text{r}}^{\text{A}}(t) \rangle_{i}\otimes\vert\psi_{\text{r}}^{\text{B}}(t)\rangle_{i},
	\label{eq-fai-sd}
\end{eqnarray}
For the equations and below, we assume to the summations to be over all non-zero Schmidt coefficients.

Here, we assume the following orthogonal conditions
\begin{equation}
	\begin{aligned}
		{\ }_{i}\langle \psi^{\text{A}}(t) \vert \psi_{\text{r}}^{\text{A}}(t)\rangle_{i'} \sim 0, \ \ 
		{\ }_{i}\langle \psi^{\text{B}}(t) \vert \psi_{\text{r}}^{\text{B}}(t)\rangle_{i'} \sim 0. 
		\label{eq-fai-orthogonal}
	\end{aligned}
\end{equation}
\R{Again, these orthogonal conditions will be asymptotically satisfied in the large-$N$ limit, meaning the accuracy of the analytical prediction will increase with $N$.} Then the entanglement spectrum of $\vert\psi'(t)\rangle$ would satisfy
\begin{equation}
	\begin{aligned}
		\boldsymbol{\Lambda}'(t) = [\Lambda_{1}(t) \cos \mu, \ \Lambda_{2}(t) \cos \mu, \cdots, \tilde{\Lambda}_{1}(t) \sin \mu, \ \tilde{\Lambda}_{2}(t) \sin \mu, \cdots]
		\label{eq-fai0-es}
	\end{aligned}
\end{equation}
Consequently, the EE of $\vert\psi'(t)\rangle$ obeys [also see Eq.~(9) in the main text]
\begin{equation}
	\begin{aligned}
		S'(t) &= -\sin^2\mu \log_{2} \sin^2\mu \sum_{i}\tilde{\Lambda}_{i} (t)^{2} -\cos^2\mu \log_{2}\cos^2\mu \sum_{i}\Lambda_{i}(t)^{2} -\sin^2\mu \sum_{i} \tilde{\Lambda}_{i}(t)^{2} \log_{2} \tilde{\Lambda}_{i} (t)^{2} - \cos^2\mu \sum_{i} \Lambda_{i}(t)^{2} \log_{2} \Lambda_{i}(t)^{2} \\ 
		&= \tilde{S}_{\mu} + \tilde{S}_{\text{P}} \sin^{2} \mu + S(t) \cos^{2} \mu.
		\label{eq-EEtime}
	\end{aligned}
\end{equation}
To obtain the last line in the above equation, we have used $\sum_{i}\tilde{\Lambda}_{i}(t)^{2} = \sum_{i}\Lambda_{i}(t)^{2} = 1$, Eq.~(\ref{eq-fai0-EEmu}), and $S(t) = -\sum_{i} \Lambda_{i}(t)^{2} \log_{2} \Lambda_{i}(t)^{2}$ that is the EE obtained by VEEF with the ``correct'' initial state. We have also used $-\sum_{i} [\tilde{\Lambda}_{i}(t)]^{2} \log_{2} [\tilde{\Lambda}_{i}(t)]^{2} = \tilde{S}_{P}$ according to the endomorphism of the VEEF time evolution for RPS's. The orthogonal conditions given by Eq.~(\ref{eq-fai-orthogonal}) concern the Schmidt basis states of a RPS and an entangled state, which are thus more difficult to satisfy than the conditions for the initial product state [Eq.~(\ref{eq-fai0-or})]. Therefore, much larger sizes are required to suppress the differences between the numerical results and the theoretical predictions (see the data in Fig.~5 in the main text).

\begin{figure}[tbp]
	\centering
	\includegraphics[width=9cm]{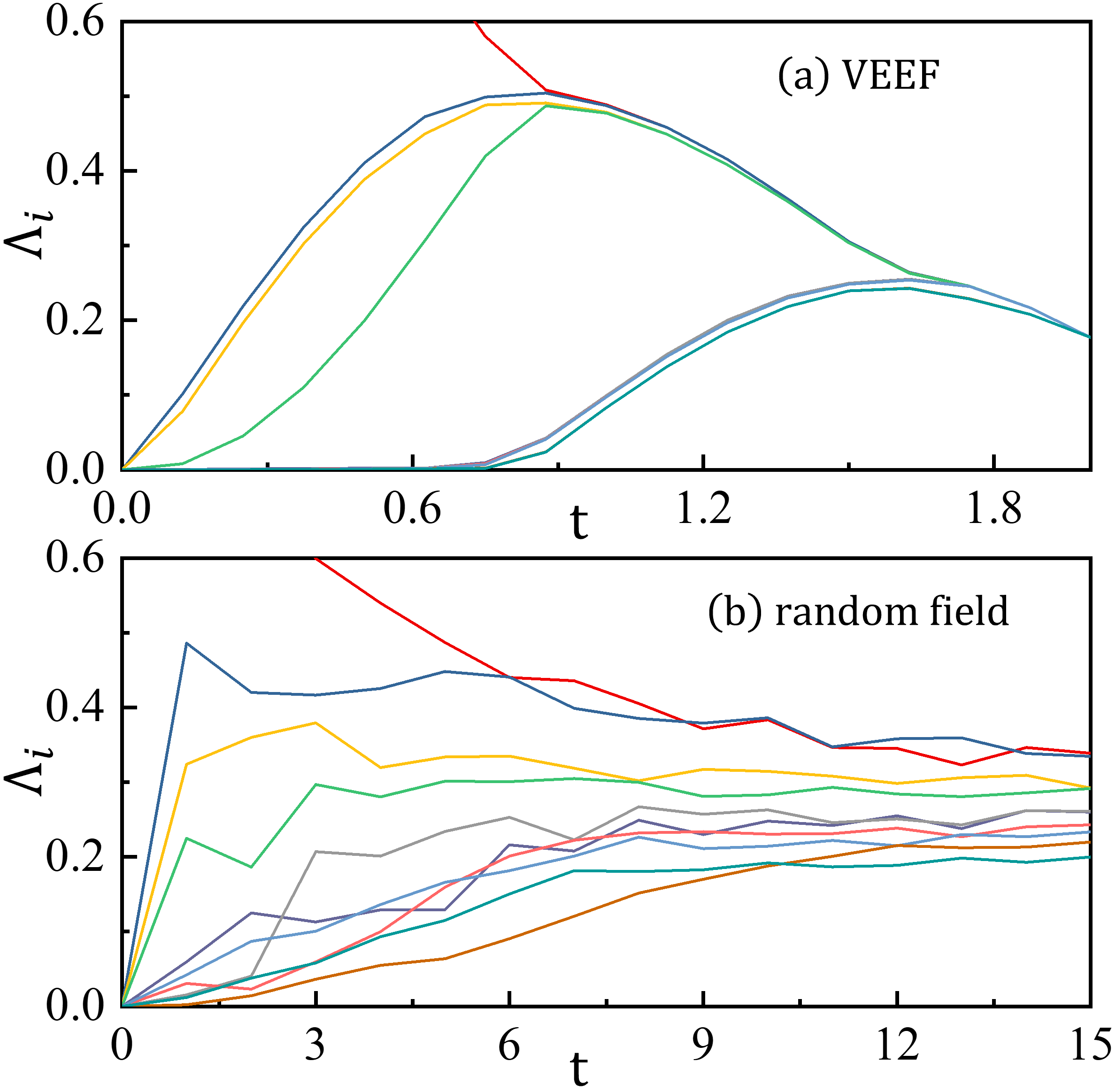}
	\caption{(Color online) The 10 largest Schmidt coefficients $\{\Lambda_{i}\}$ ($i=1,2,\cdots,10$) of $\vert \psi(t) \rangle$ with the evolution under (a) VEEF for $t \leq T_{S}$ and (b) random field for $t \leq T$ with $T \gg T_{S}$. We consider the $N=10$ quantum Ising chain with periodic boundary condition.}
	\label{fig-spc}
\end{figure}

The Schmidt coefficients of the states with $S=\tilde{S}$ and those with $S=\tilde{S}_{\text{P}}$~\cite{PhysRevB.100.125115} exhibit obvious difference. Fig.~\ref{fig-spc} shows the 10 largest Schmidt coefficients $\{\Lambda_{i}\}$ ($i=1,2,\cdots,10$) of the state $\vert \psi(t) \rangle$ with the evolution under VEEF and random field. We consider the $N=10$ quantum Ising chain with periodic boundary condition. In (a) with VEEF, the Schmidt coefficients converge approximately to the same value  with $\Lambda_{i} = \sqrt{2^{-5}} \simeq 0.177$ for $t \to T_{S}$ (with $S(t) \to \tilde{S}$). In (b) with a random field, the Schmidt coefficients range from $0.2$ from $0.34$ for $t \gg 10$ (with $S(t) \to \tilde{S}_{\text{P}}$). Note the initial state is taken as a product state, and thus we have $\Lambda_{1} = 1$ and $\Lambda_{2} =  \Lambda_{3} = \cdots = 0$ at $t=0$.

\R{Such a quantum state with equal Schmidt coefficients can be regarded as the Choi state of a unitary operations~\cite{PhysRevA.101.052311}. The Choi states play fundamental roles in various quantum information processing tasks such as gate teleportation, quantum process tomography, and measurement-based quantum computation~\cite{PhysRevA.101.052311, PRXQuantum.3.020344, PhysRevA.107.042403}.}

\R{To be more specific, the Choi state of an operator, say $\hat{U} = \sum_{ss'} U_{ss'} |s\rangle \langle s'|$, is obtained by identically transforming the \textit{bra} basis $\langle s'|$ to \textit{ket} basis $|s' \rangle$, as $|U\rangle = \frac{1}{Z} \sum_{ss'} U_{ss'} |s\rangle |s'\rangle $ (with $Z$ a normalization factor). The Schmidt coefficients of the state $|U\rangle$ are equal when the operator $\hat{U}$ is unitary, and \textit{vice versa}. Note $s$ (or $s'$ with $\dim(s)=\dim(s')$) may denote one spin or a subsystem containing several spins.}

\R{Based on the above relation between $\hat{U}$ and its Choi state $|U\rangle$, various quantum computational tasks can be realized by $|U\rangle$. One example is the quantum process tomography, where one can implement quantum state tomography on $|U\rangle$ to reconstruct $\hat{U}$. Another example is the measurement-based quantum computation, in which the unitary operations are realized by measurements on the Choi state. To realize the operation $\hat{U} |\psi\rangle = |\phi \rangle$, one may construct projective measurement on the $s'$ part of $|U\rangle$ and ensure the projection results in $|\psi\rangle$ by post selection. For simplicity, we assume the coefficients of $|\psi\rangle$ to be real, otherwise one should carefully deal with the signs of the imaginary parts. The state after the measurement gives the one after the operation as $|\phi\rangle = \hat{U} |\psi\rangle = \frac{1}{Z} \langle \psi|U\rangle$.}

% \R{The importance of Choi states is evident in their utility for representing and implementing quantum operations or channels in real-world scenarios such as gate teleportation and stored-program quantum computing. In gate teleportation, Choi states facilitate the teleportation of quantum gates across different locations, while in stored-program quantum computing, they are essential for storing and executing quantum programs effectively.}

% \R{The recognition of the maximally entangled state corresponding to a flat entanglement spectrum as a Choi state underscores its significance in understanding and analyzing quantum systems. By leveraging Choi states, researchers can gain insights into the properties of quantum channels, enhancing the efficiency of tasks like quantum entanglement transfer and quantum state synthesis. This highlights the importance of identifying the maximally entangled state in practical quantum information processing scenarios.}

\section{III. Supplemental data on the robustness of ballistic spreading}

\begin{figure}[tbp]
	\centering
	\includegraphics[width=8cm]{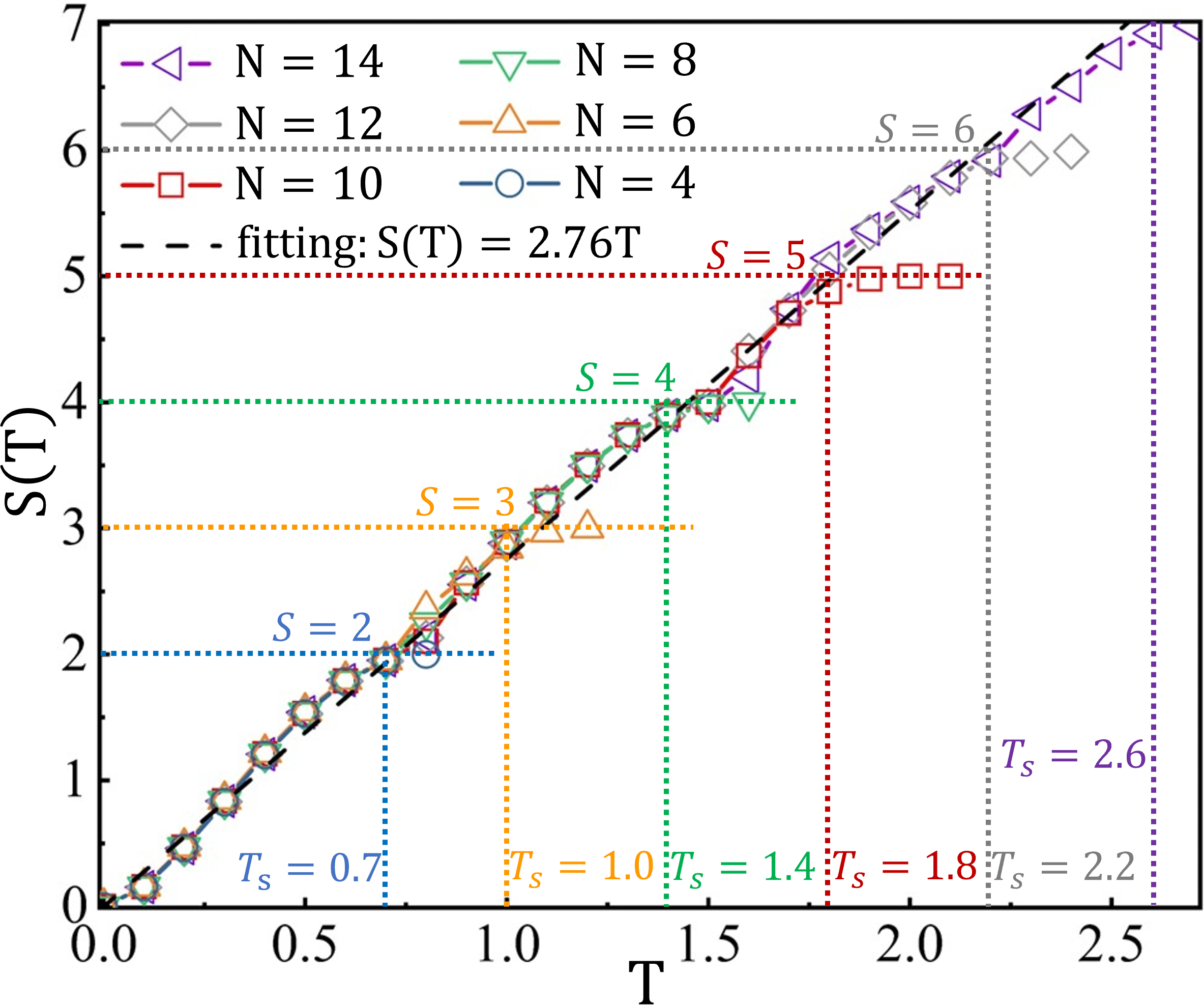}
	\caption{(Color online) The EE $S(T)$ of the final state versus the total evolution time $T$ for different numbers of spins $N$ in the quantum Ising chain. In all cases, the EE increases following the linear relation $S(T)=vT$ for $T\leq T_{S}$, and converges to the genuine saturation $\tilde{S} = N/2$. For all sizes ($N=4, \cdots, 14$), we have $v\simeq 2.76$. Here, we consider the quantum Ising chain.}
	\label{fig3}
\end{figure}

\begin{figure}[t]
	\centering
	\includegraphics[width=8cm]{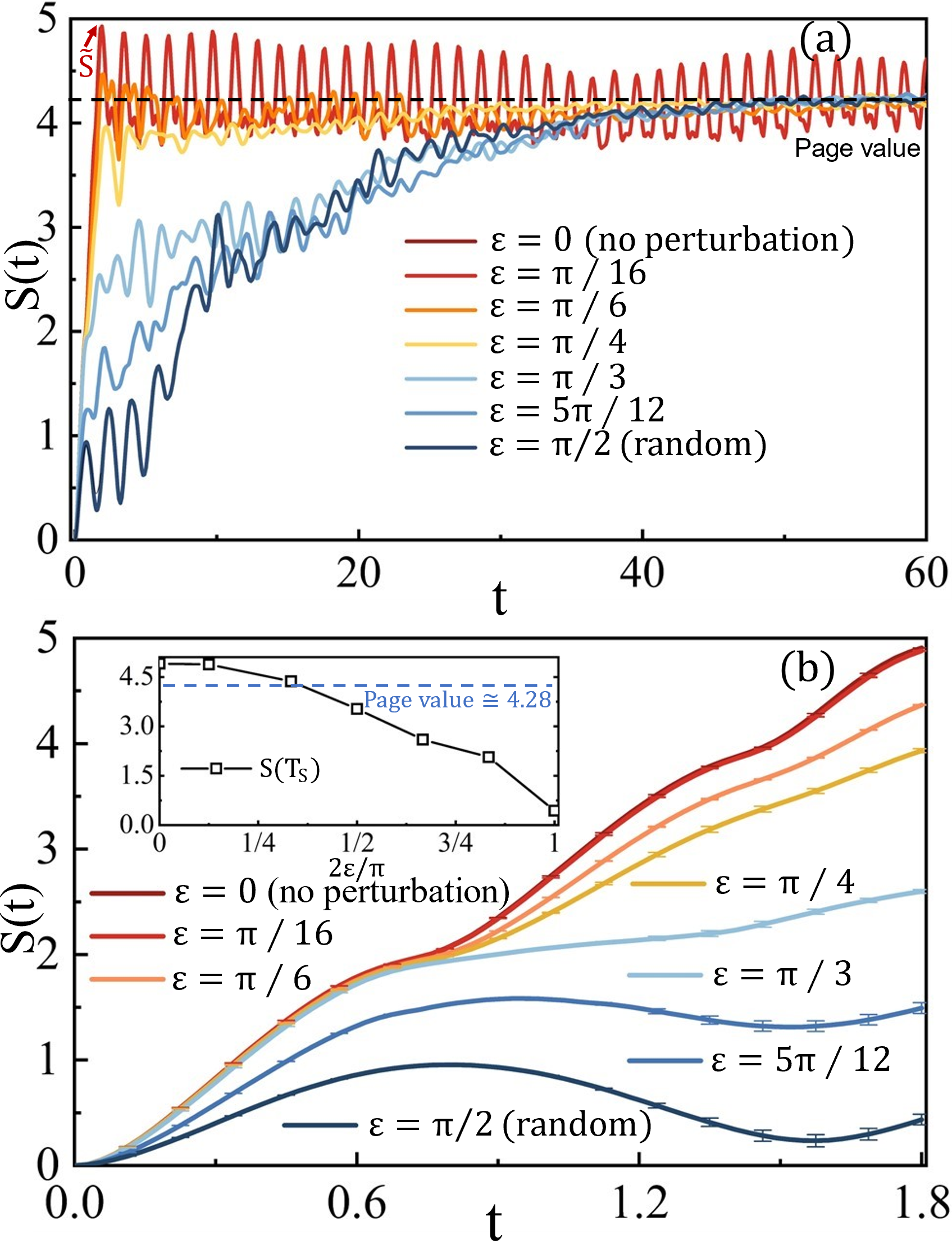}
	\caption{(Color online) The EE $S(t)$ against the time $t$ with different $\varepsilon$ perturbing the VEEF [see Eq.~(\ref{eq-h1})] (a) for $t\leq 60$, and (b) for $t \leq T_S=5$. The inset of (b) shows the EE of the final state ($S(T)$) versus $\varepsilon$. Here, we consider the $N=10$ quantum Ising chain.}
	\label{fig-3}
\end{figure}

The ballistic spreading is robust for different system sizes. Fig.~\ref{fig3} shows the EE $S(T)$ of the final state against the total evolution time $T$. As an example, we consider the one-dimensional (1D) quantum Ising model (QIM) with periodic boundary condition, with its Hamiltonian $ \hat{H}_{\text{QIM}}(t) =  \sum_{n=1}^{N-1} \hat{\sigma}_n^{z} \hat{\sigma}_{n+1}^{z} + \hat{\sigma}_1^{z} \hat{\sigma}_{N}^{z} + \sum_{n=1}^N \sum_{\alpha=x,z} h^{\alpha}_n(t) \hat{\sigma}_n^{\alpha}$. For $T \geq T_{S}=\frac{N}{2v}$, the EE reaches the maximum with $S(T)=\tilde{S}=N/2$ (marked by the vertical and horizontal dot lines). For $T < T_{S}$, we have $S(T)=vT$ that is the maximal EE reachable in the evolution time $T$ with the velocity $v$. This is consistent with the linear growth of $S(t)$ (with $0 \leq t \leq T$) during the time evolution. We robustly have $v \simeq 2.76$ for the QIM of different sizes ($N=4, \cdots, 14$).

Fig.~\ref{fig-3} illustrates the robustness against random perturbations on the VEEF. The time-dependent field is perturbed as
\begin{equation}
	h'^{\alpha}_n(t) =  \left\{
	\begin{aligned}
		& h^{\alpha}_n(t) \cos\varepsilon+ 
		\tilde{h}^{\alpha}_n(t) \sin\varepsilon \ \ \text{for } t \leq T_S, \nonumber \\ 
		& \tilde{h}^{\alpha}_n(t) \sin\varepsilon \ \ \ \ \ \ \ \ \ \ \ \ \ \ \ \ \ \ \ \text{for } t > T_S,
	\end{aligned}
	\right.
	\label{eq-h1}
\end{equation}
with $\varepsilon \in [0, \pi/2]$ controlling the strength of perturbation.

In Fig.~\ref{fig-3}(a), the curve for $\varepsilon = \pi/2$ demonstrates the EE with a random field $\tilde{h}^{\alpha}_n(t)$. The $S'(t)$ approaches to the Page value as expected. For $0< \varepsilon  < \pi/2$, the Page value is approached in a faster speed than that with a random field, even though the VEEF is implemented only during a short time duration $t \leq T_{S}$ [see Eq.~(\ref{eq-h1})]. For $\varepsilon = 0$, no random perturbation is added, and $S'(t)$ reaches the maximum $\tilde{S}$ at $t=T_S$ as expected. For $t>T_S$, $S'(t)$ oscillates around the Page value, as the evolution is just given by the coupling terms (with zero fields).

In Fig.~\ref{fig-3}(b), the area for $t \leq T_{S}$ is zoomed in. As $\varepsilon$ increases, the growth of EE progressively deviated from being linear to $t$. The inset shows the EE $S(T_{S})$ for different $\varepsilon$. We have $S(T_{S}) > \tilde{S}_P$ for $\varepsilon < \pi/6$.

\begin{figure}[tbp]
	\centering
	\includegraphics[width=9cm]{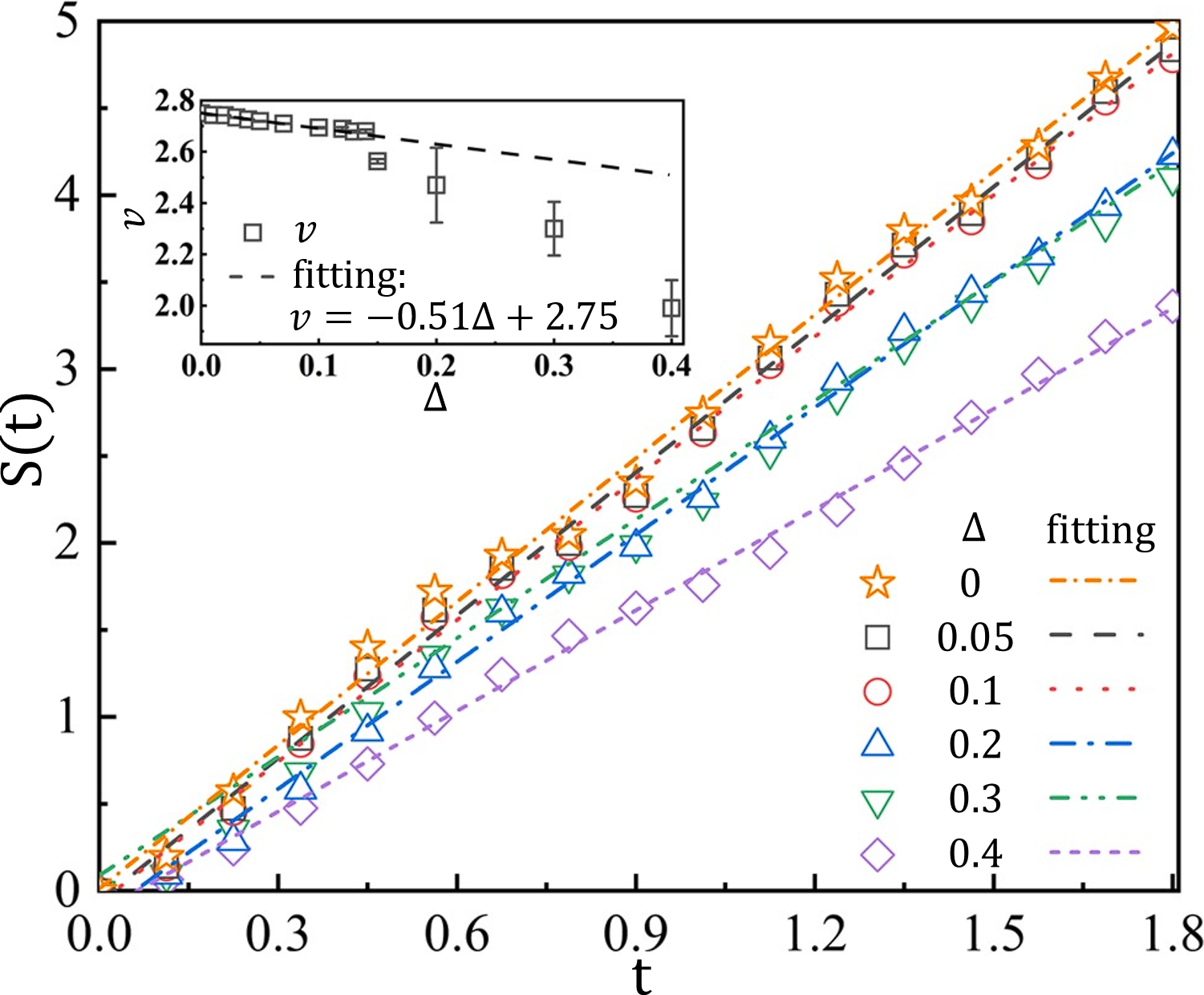}
	\caption{\R{(Color online) The EE $S(t)$ of the quantum Ising chain under VEEF by randomly perturbing the coupling strengths from uniformly $1$ to $(1+\tilde{J}_{n})$, where $n$ denotes the position of the coupling and $\tilde{J}_{n}$ is a random number generated by the Gaussian distribution taking its mean as zero and its standard deviation as $\Delta$. The ballistic spreading behavior is robust under such random perturbations with different values of $\Delta$. The inset shows that the velocity $v$ decreases linearly with $\Delta$ for about $\Delta < 0.14$. The error bars show the standard deviations of 10 independent simulations.}}
	\label{fig-PertCoup}
\end{figure}

\R{The ballistic spreading is also robust to the random perturbations on the coupling strengths. We add randomness on the two-body terms, where the coupling in $n$-th local Hamiltonian in the chain is perturbed to 
	\begin{equation}
		\hat{H}_{n,n+1} = (1+\tilde{J}_{n}) \hat{\sigma}_{n}^{z} \hat{\sigma}_{n+1}^{z},
		\label{eq-RandCoup}
	\end{equation}
	where $\tilde{J}_{n}$ is randomly generated by the Gaussian distribution with standard deviation $\Delta$ (we take its mean as zero). Thus, $\Delta$ characterizes the strengths of the randomness. Fig.~\ref{fig-PertCoup} shows that the ballistic EE spreading behavior is robust to the random perturbations on the coupling strengths, in which the velocity is lowered as $\Delta$ increases. The inset shows that for about $\Delta <0.14$, $v$ exhibits linear dependence on $\Delta$ as $v = -0.51 \Delta + 2.75$.}

\begin{figure}[tbp]
	\centering
	\includegraphics[width=9cm]{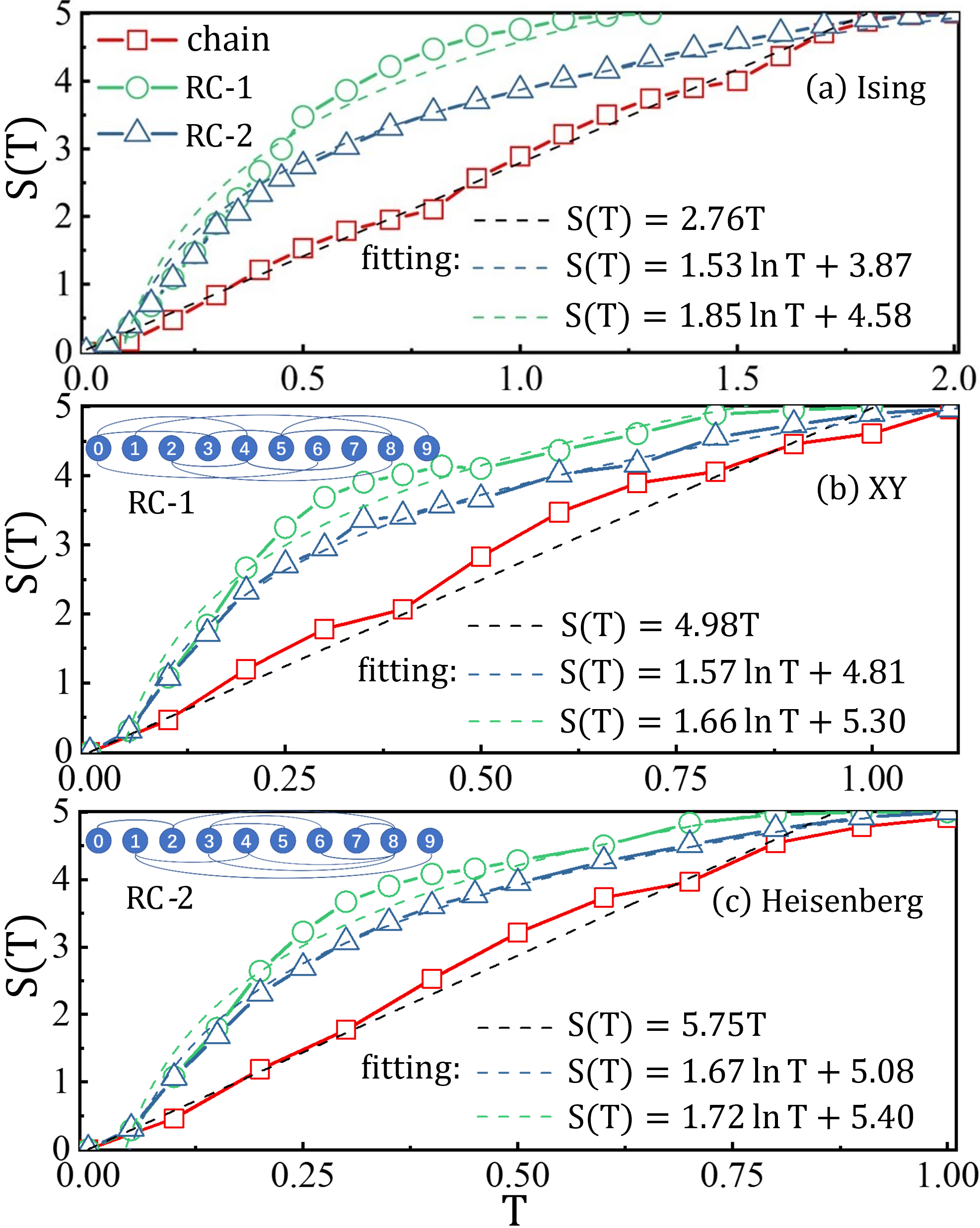}
	\caption{(Color online) The EE $S(T)$ of the final state versus the total evolution time $T$ with (a) Ising, (b) XY, and (c) Heisenberg interactions. Linear growth of $S(T)$ is observed for the 1D spin chains (red squares), with the velocity $v \simeq 2.76$, $4.98$, and $5.75$ for the Ising, XY, and Heisenberg interactions, respectively. For the lattices with random connections (RC), which are denoted as RC-1 and RC-2 [see the insets of (b) and (c), respectively], faster growths of EE are demonstrated.}
	\label{fig-1}
\end{figure}

The spreading of EE is known to be relevant to the geometry of lattice ~\cite{anshu_entanglement_2022, rico_ortega_entanglement_2022}. The growth of EE should not be faster than being linear when the lattice geometry obeys the 1D area law~\cite{PhysRevB.95.094302}. Fig.~\ref{fig-1} shows that long-range interactions breaking the 1D area law might lead to a super-ballistic EE growth. For the 1D chains (with nearest-neighboring couplings and periodic boundary condition), the $S(T)$ grows linearly with the Ising ($\hat{H}_{n,n+1} = \sum_{\alpha=z} \hat{\sigma}_n^{\alpha} \hat{\sigma}_{n+1}^{\alpha} $), XY ($\hat{H}_{n,n+1} = \sum_{\alpha=x,y} \hat{\sigma}_n^{\alpha} \hat{\sigma}_{n+1}^{\alpha} $), and Heisenberg ($\hat{H}_{n,n+1} = \sum_{\alpha=x,y,z} \hat{\sigma}_n^{\alpha} \hat{\sigma}_{n+1}^{\alpha} $) interactions with the velocity $v=2.76$, $4.98$, and $5.75$, respectively. The linear growth can be exceeded with the presence of long-range interactions. We demonstrate this by two lattices with random connections (RC) [see the lattice geometries in the insets of Fig.~\ref{fig-1} (b) and (c), which are denoted as RC-1 and RC-2]. The probability of coupling each two spins equals to $p = \frac{N}{N!}$. The strength of each spin coupling is uniformly assigned as $1$. The whole $S(T)$ curves of the random lattices are beyond the curve of the linear growth, implying faster growth of EE with the long-range interactions. The main bodies of $S(T)$ curves can be fitted by the logarithmic functions. The EE of the final state still converges to the maximum $\tilde{S}$, since $\tilde{S}$ is solely determined by the degrees of freedom in the subsystems and is irrelevant to the interactions (as long as the lattice contains no disconnected part). 

\R{Our results with different kinds of perturbations uncover the relations between the EE spreading and Hamiltonian, which is a critical issue about quantum dynamics. We may give the following theoretical implications:}

\begin{itemize}
	\item \R{For a quantum Hamiltonian consisting of one- and two-body terms, the spreading behavior of EE is maximally ballistic with respect to time $t$ when the lattice geometry (determined by the two-body terms) is 1D.}
	\item \R{The above maximal growth of EE can be reached by tuning the one-body terms of the Hamiltonian.}
\end{itemize}

\R{We regard the above two as conjectures on the controllability of EE spreading by adjusting the one-body terms of Hamiltonian. While keeping the 1D lattice geometry unchanged, our numerical simulations of VEEF show that the ballistic spreading of EE is robustly obtained by variationally adjusting the magnetic field (formed by one-body terms). The maximal velocity of EE spreading under VEEF depends mainly on the two-body terms, including their types, strengths, and the lattice geometry given by these terms, while it is independent on the system size and total evolution duration (in the cases of $T \leq T_{\text{S}}$). Be noted that the spreading behavior of EE is robustly ballistic under VEEF when the perturbations do not change the 1D geometry, but will become super-ballistic with the presence of long-range interactions.}

Besides EE, another important quantity to reveal the properties of quantum dynamics is R$\acute{e}$nyi entropy (RE)~\cite{PhysRevLett.122.250602, Huang_2020}. The RE of $\vert \psi(t) \rangle$ is defined as
\begin{equation}
	S_\alpha(t) \equiv \frac{1}{1-\alpha} \log_{2} \text{Tr}_{\text{A}}\left[\hat{\rho}(t)^{\alpha}\right],
	\label{eq-renyi}
\end{equation}
with $\hat{\rho}(t)=\mathrm{Tr}_B|\psi(t)\rangle\langle\psi(t)|$ the reduced density matrix of a sub-system $A$ by tracing over the degrees of freedom of the rest (denoted as B). We have $S_\alpha(t) \to S(t)$ for $\alpha \to 1$. 

\begin{figure}[tbp]
	\centering
	\includegraphics[width=15cm]{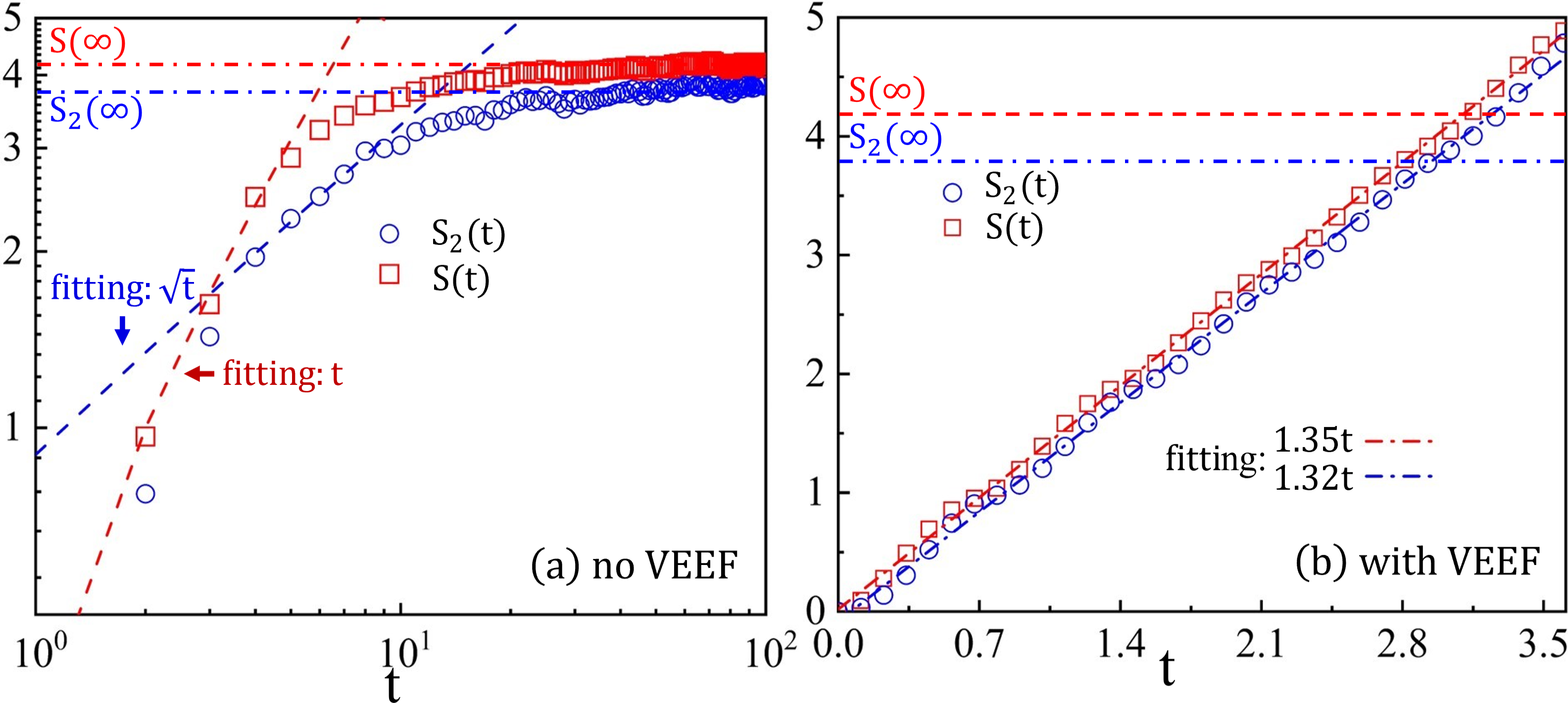}
	\caption{(Color online) (a) Without VEEF (fixing $h^x_n(t)=0.9045$ and $h^z_n(t)=0.8090$), the EE $S(t)$ and RE $S_2(t)$ exhibits the ballistic and sub-ballistic growths for $t \sim O(1)$, respectively~\cite{PhysRevLett.122.250602}. The convergent EE and RE in the long-time limit are denoted as $S(\infty)$ and $S_2(\infty)$, respectively. (b) With the VEEF that is optimized to maximize EE, both the EE and RE exhibit persistent ballistic spreading, and both values at $t=T_{S}$ exceed the convergent values $S(\infty)$ and $S_2(\infty)$ in (a). Here, we consider the $N=10$ quantum Ising chain with open boundary condition.}
	\label{fig-Renyi}
\end{figure}

We show that the VEEF by maximizing the EE will also lead to the ballistic spreading of RE. Fig.~\ref{fig-Renyi} shows the EE $S(t)$ and the RE $S_{2}(t)$ (taking $\alpha=2$) with or without VEEF, on quantum Ising chain with open boundary condition. In the duration of $t \sim O(1)$, the EE and RE grow in different manners by fixing $h^x_n(t)=0.9045$ and $h^z_n(t)=0.8090$~\cite{PhysRevLett.122.250602}. The EE obeys ballistic spreading with $S(t)\propto t$. The RE obeys sub-ballistic spreading, with $S_{2}(t)\propto \sqrt{t}$ before the convergence is approached. This is consistent with the previous results~\cite{PhysRevLett.122.250602}. In Fig.~\ref{fig-Renyi}(b) with the VEEF maximizing EE, we show that both EE and RE exceed the convergent values $S(\infty)$ and $S_2(\infty)$ shown in (a), and they both increase linearly for $t \leq T_{S}$. 

\section{IV. VEEF data for quantum spin chains}

\begin{figure}[htbp]
	\centering
	\includegraphics[width=19cm]{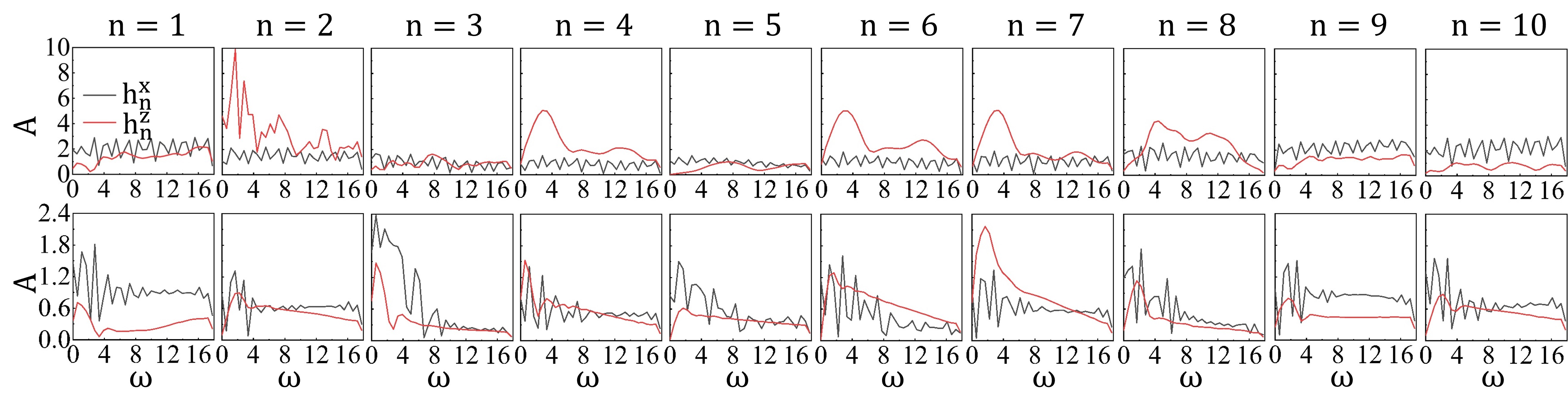}
	\caption{\R{(Color online) The amplitude $A$ with different frequency $\omega$, obtained by the FFT on $h_n^x(t)$ and $h_n^z(t)$ for the quantum Ising chain. The two rows show the results before and after adding the regularization term, respectively.}}
	\label{fig-IsingVEEF}
\end{figure}

\begin{figure}[htbp]
	\centering
	\includegraphics[width=19cm]{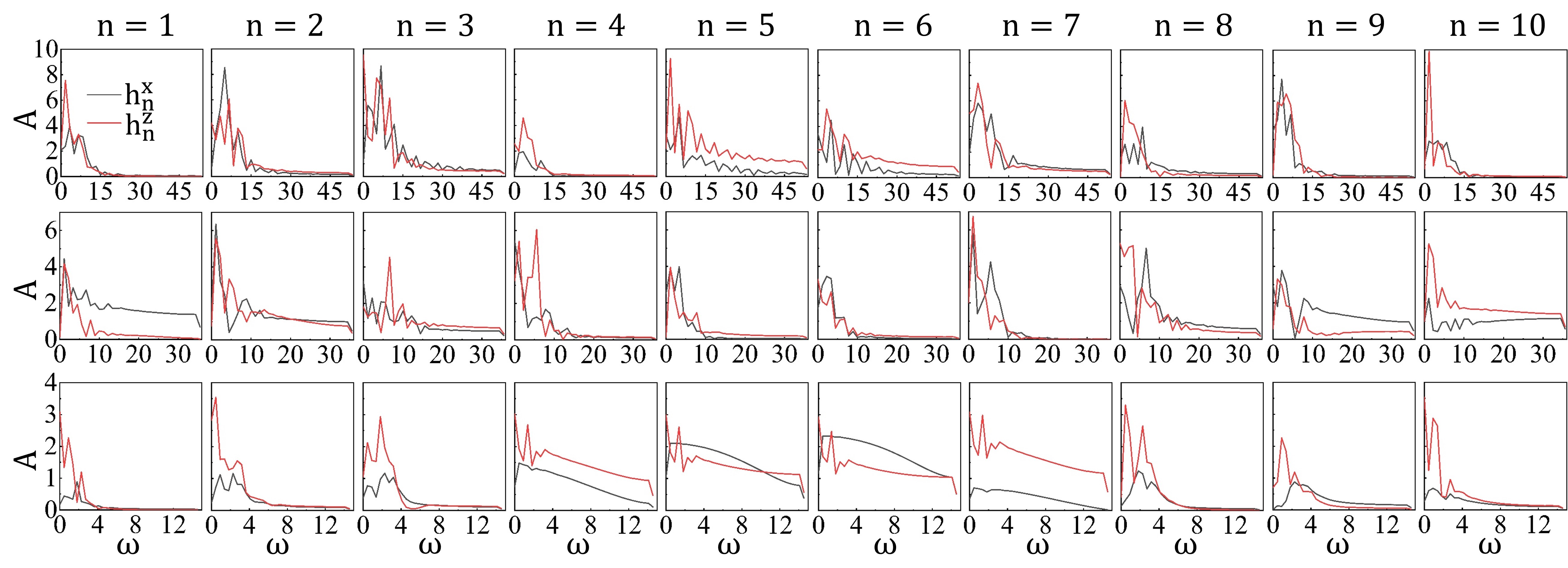}
	\caption{\R{(Color online) The amplitude $A$ with different frequency $\omega$, obtained by the FFT on $h_n^x(t)$ and $h_n^z(t)$. The three rows show the results for the XXZ, Heisenberg, and XY chains, respectively, after adding the regularization term.}}
	\label{fig-ChainVEEF}
\end{figure}

Below, we implement the fast Fourier transform (FFT) on the VEE for the $N = 10$ quantum Ising chain with periodic boundary condition. The FFT satisfies
\begin{equation}
	\begin{aligned}
		\R{h(t)=\sum_{k} A_{k} e^{-i \omega_{k} t},}
		\label{eq-FFT}
	\end{aligned}
\end{equation}
\R{with $A_{k}$ is the amplitude and $\omega_{k}$ the frequency.}

\R{Meanwhile, we consider to add a regularization term to the optimization problem, which becomes 
	\begin{equation}
		\begin{aligned}
			\max \left( S(T) - \lambda \sqrt{\sum_{\alpha, n, t} |h^{\alpha}_n(t)|^2} \right),
			\label{eq-RTmax}
		\end{aligned}
	\end{equation}
	with $\lambda$ a small positive constant. Note $t$ is discretized by Trotter decomposition. This added term, which is also known as the Lagrangian term in the field of numerical optimization, is to get the maximal EE with the magnetic field being as small as possible.}

\R{Fig.~\ref{fig-IsingVEEF} shows the FFT results of VEEF for the quantum Ising chain with $N=10$ spins before (first row) and after (second row) adding the regularization term. The regularization term largely suppresses the amplitudes while keeping the EE almost unchanged. The maximal amplitude is reduced approximately from $10$ to $1.8$ and the amplitudes generally decreases with the frequency $\omega$ after adding the regularization term. The difference of EE with or without the regularization term is just about $O(10^{-3})$. Similar observations are made for the XXZ, Heisenberg, and XY chains (see Fig.~\ref{fig-ChainVEEF}). The reduction of the amplitudes, particularly those of the high-frequency parts are in great favor of the experimental implementations.}

In Table \ref{table-1}, we provide the data of VEEF \R{after adding the regularization term} for the $N = 10$ quantum Ising chain with periodic boundary condition. 

\begin{table}[ht]
	\centering
	\caption{The data of VEEF for the $N=10$ quantum Ising chain with periodic boundary condition \R{after adding the regularization term.} We take the total evolution time $T=1.8$ and the number of Trotter slices $k$ up to 64 (meaning the Trotter step $\tau=1.8/64$).} \label{table-1}
	\begin{tabular*}{18cm}{@{\extracolsep{\fill}}lcccccccccc}
		\\
		\hline\hline
		\textbf{VEEF} & $n=1(x)$ & $n=2(x)$ & $n=3(x)$ & $n=4(x)$ & $n=5(x)$ & $n=6(x)$ & $n=7(x)$ & $n=8(x)$ & $n=9(x)$ & $n=10(x)$ \\ \hline
		k = 1 & -29.5699 & 18.9182 & -8.3312 & 14.3917 & -9.7741 & 8.7372 & 17.8110 & 13.0853 & -26.6152 & 20.9758 \\ \hline
		k = 2 & 1.7548 & 0.6794 & -5.6305 & 1.5555 & -0.8308 & 1.8709 & 2.6165 & 7.3209 & -0.3334 & -1.0327 \\ \hline
		k = 3 & -0.0523 & -0.1339 & -3.0311 & -0.1154 & 0.0514 & -0.0688 & 0.8269 & 2.6857 & 0.2813 & 0.0867 \\ \hline
		k = 4 & -0.0812 & 0.0134 & -1.6595 & 0.0320 & -0.0039 & 0.0527 & 0.1746 & 0.7983 & -0.1279 & 0.0104 \\ \hline
		k = 5 & -0.0139 & -0.0372 & -0.9606 & -0.0230 & 0.0076 & 0.0050 & 0.0997 & -0.0046 & 0.0205 & 0.0011 \\ \hline
		k = 6 & -0.0278 & -0.0242 & -0.6197 & -0.0247 & 0.0064 & 0.0076 & 0.0867 & -0.0684 & 0.0036 & 0.0060 \\ \hline
		k = 7 & -0.0223 & -0.0189 & -0.3386 & -0.0510 & 0.0028 & 0.0064 & 0.0830 & -0.1351 & 0.0088 & 0.0069 \\ \hline
		k = 8 & -0.0165 & -0.0261 & -0.1185 & -0.0724 & -0.0013 & 0.0025 & 0.0695 & -0.0089 & 0.0230 & 0.0022 \\ \hline
		k = 9 & 0.0188 & -0.0814 & 0.2602 & -0.0807 & -0.0032 & -0.0067 & 0.0458 & 0.0934 & 0.0631 & -0.0283 \\ \hline
		k = 10 & 0.1246 & -0.2199 & 0.6966 & -0.0420 & 0.0047 & -0.0232 & 0.0036 & 0.3289 & 0.1698 & -0.1262 \\ \hline
		k = 11 & 0.3247 & -0.4567 & 1.3119 & 0.0790 & 0.0302 & -0.0452 & -0.0602 & 0.6660 & 0.3566 & -0.3144 \\ \hline
		k = 12 & 0.5437 & -0.6924 & 1.7787 & 0.2619 & 0.0682 & -0.0604 & -0.1276 & 1.0063 & 0.5529 & -0.5233 \\ \hline
		k = 13 & 0.7018 & -0.8437 & 1.9845 & 0.4301 & 0.0982 & -0.0672 & -0.2018 & 1.3620 & 0.7138 & -0.6813 \\ \hline
		k = 14 & 0.7380 & -0.8309 & 1.6474 & 0.4857 & 0.1006 & -0.0561 & -0.2285 & 1.4015 & 0.7430 & -0.7238 \\ \hline
		k = 15 & 0.7410 & -0.7323 & 0.9506 & 0.4145 & 0.0701 & -0.0513 & -0.2451 & 1.3334 & 0.7387 & -0.7399 \\ \hline
		k = 16 & 0.7104 & -0.5674 & 0.0394 & 0.2725 & 0.0253 & -0.0488 & -0.2242 & 1.0351 & 0.6750 & -0.7218 \\ \hline
		k = 17 & 0.7019 & -0.4102 & -0.7953 & 0.1160 & -0.0307 & -0.0625 & -0.1884 & 0.5911 & 0.6049 & -0.7230 \\ \hline
		k = 18 & 0.7397 & -0.2770 & -1.4853 & -0.0169 & -0.0794 & -0.0839 & -0.1690 & 0.3400 & 0.5936 & -0.7750 \\ \hline
		k = 19 & 0.7762 & -0.1591 & -2.0095 & -0.1771 & -0.1870 & -0.1674 & -0.1711 & -0.0975 & 0.5506 & -0.8196 \\ \hline
		k = 20 & 0.9761 & -0.1416 & -2.4059 & -0.4790 & -0.6376 & -0.5990 & -0.3869 & -0.2374 & 0.6514 & -1.0355 \\ \hline
		k = 21 & 0.9008 & -0.0348 & -2.5786 & -0.8256 & -1.3175 & -1.2516 & -0.6113 & -0.9106 & 0.5016 & -0.9535 \\ \hline
		k = 22 & 0.7627 & 0.0747 & -2.3812 & -1.0081 & -1.9421 & -1.8511 & -0.7049 & -1.8203 & 0.2334 & -0.7922 \\ \hline
		k = 23 & -0.9988 & 0.9807 & -1.6694 & 0.1924 & 0.4400 & 0.5068 & 0.6507 & -3.4543 & -1.0938 & 1.0401 \\ \hline
		k = 24 & -3.8674 & 2.2811 & -0.7061 & 2.0201 & 4.5274 & 4.5286 & 2.4905 & -4.6239 & -2.7619 & 3.9676 \\ \hline
		k = 25 & -7.3343 & 3.9643 & 0.2485 & 3.9425 & 7.8926 & 7.8419 & 4.3277 & -4.4146 & -4.6094 & 7.4637 \\ \hline
		k = 26 & -6.7768 & 4.5338 & 1.0144 & 4.4239 & 6.0203 & 6.0103 & 4.6233 & -2.5623 & -5.0568 & 6.8881 \\ \hline
		k = 27 & -4.6855 & 4.3741 & 1.6424 & 4.0779 & 3.7747 & 3.8102 & 4.1316 & -0.3871 & -4.5823 & 4.7378 \\ \hline
		k = 28 & -3.4743 & 4.1425 & 2.2334 & 4.0061 & 3.0417 & 3.0921 & 4.0055 & 0.6828 & -4.0205 & 3.4494 \\ \hline
		k = 29 & -2.6516 & 3.7459 & 2.7671 & 3.2427 & 2.1745 & 2.2515 & 3.3149 & 0.8128 & -3.2944 & 2.5414 \\ \hline
		k = 30 & -2.1657 & 3.3738 & 3.2582 & 2.3310 & 1.5302 & 1.6339 & 2.5440 & 0.6888 & -2.6986 & 2.0035 \\ \hline
		\\ 
	\end{tabular*}
\end{table}

\begin{table}[tbp]
	\centering
	\scalebox{0.99}{
		\begin{tabular*}{18cm}{@{\extracolsep{\fill}}lcccccccccccc}
			\\
			\hline\hline
			\textbf{VEEF} & $n=1(x)$ & $n=2(x)$ & $n=3(x)$ & $n=4(x)$ & $n=5(x)$ & $n=6(x)$ & $n=7(x)$ & $n=8(x)$ & $n=9(x)$ & $n=10(x)$ \\ \hline
			k = 31 & -1.1332 & 2.0626 & 3.4025 & 1.0232 & 0.6543 & 0.7079 & 1.1038 & 1.3403 & -1.3212 & 0.9758 \\ \hline
			k = 32 & -0.5992 & 1.1836 & 2.9012 & 0.4353 & 0.3020 & 0.3012 & 0.2996 & 2.4418 & -0.4883 & 0.4720 \\ \hline
			k = 33 & 0.1672 & -0.1147 & 1.3279 & 0.3444 & 0.1392 & 0.0449 & -0.2431 & 3.5321 & 0.4311 & -0.1937 \\ \hline
			k = 34 & 0.4454 & -0.9171 & -1.4029 & 1.0061 & 0.4006 & 0.2519 & 0.0041 & 3.8220 & 0.5473 & -0.3410 \\ \hline
			k = 35 & 0.6636 & -1.6378 & -4.8832 & 1.9405 & 0.6876 & 0.4775 & 0.3641 & 3.5196 & 0.4695 & -0.4406 \\ \hline
			k = 36 & 0.5040 & -1.5901 & -7.3007 & 2.5142 & 0.8441 & 0.6490 & 0.6735 & 3.0181 & 0.1266 & -0.2807 \\ \hline
			k = 37 & 0.1053 & -0.7696 & -7.1019 & 1.9911 & 0.6160 & 0.5445 & 0.6519 & 2.6417 & -0.2528 & 0.0094 \\ \hline
			k = 38 & -0.2316 & 0.1043 & -5.4389 & 0.9917 & 0.2989 & 0.3653 & 0.5186 & 2.4421 & -0.5154 & 0.2422 \\ \hline
			k = 39 & -0.4807 & 0.6842 & -3.9845 & 0.1768 & 0.0126 & 0.1628 & 0.3749 & 2.3560 & -0.6986 & 0.4346 \\ \hline
			k = 40 & -0.5401 & 0.8866 & -3.2071 & -0.2326 & -0.1034 & 0.0709 & 0.3399 & 2.3232 & -0.7459 & 0.4806 \\ \hline
			k = 41 & -0.5776 & 0.9660 & -2.8780 & -0.4630 & -0.1943 & -0.0205 & 0.3379 & 2.3173 & -0.7662 & 0.5181 \\ \hline
			k = 42 & -0.5647 & 0.9617 & -2.7775 & -0.5760 & -0.2045 & -0.0473 & 0.3858 & 2.3329 & -0.7537 & 0.5129 \\ \hline
			k = 43 & -0.5694 & 0.9476 & -2.7662 & -0.6990 & -0.2668 & -0.1303 & 0.4070 & 2.3756 & -0.7443 & 0.5255 \\ \hline
			k = 44 & -0.5677 & 0.9186 & -2.7804 & -0.8425 & -0.3658 & -0.2541 & 0.4051 & 2.4366 & -0.7328 & 0.5326 \\ \hline
			k = 45 & -0.6298 & 0.9014 & -2.7820 & -1.0788 & -0.6588 & -0.5730 & 0.3196 & 2.4871 & -0.7371 & 0.6016 \\ \hline
			k = 46 & -0.6738 & 0.8769 & -2.7427 & -1.3759 & -1.3800 & -1.3194 & 0.1640 & 2.4727 & -0.7414 & 0.6534 \\ \hline
			k = 47 & -1.0840 & 0.9379 & -2.6020 & -1.3024 & -1.3476 & -1.3334 & 0.3004 & 2.3322 & -0.8215 & 1.0654 \\ \hline
			k = 48 & -1.5848 & 1.0005 & -2.2808 & -0.8223 & -0.2174 & -0.2498 & 0.7760 & 2.0446 & -0.9016 & 1.5707 \\ \hline
			k = 49 & -2.8440 & 1.1341 & -1.8005 & 0.2120 & 5.2281 & 5.1107 & 1.5606 & 1.6346 & -1.0365 & 2.8265 \\ \hline
			k = 50 & -3.4085 & 1.1062 & -1.3153 & 0.6978 & 8.0695 & 7.9043 & 1.5074 & 1.2050 & -1.0164 & 3.3978 \\ \hline
			k = 51 & -3.4656 & 0.9936 & -0.9297 & 0.5600 & 5.8983 & 5.7600 & 1.0886 & 0.8376 & -0.9113 & 3.4621 \\ \hline
			k = 52 & -3.0901 & 0.8485 & -0.6448 & 0.1879 & 3.4332 & 3.3538 & 0.8007 & 0.5525 & -0.7721 & 3.0953 \\ \hline
			k = 53 & -2.6993 & 0.7176 & -0.4318 & -0.0751 & 2.4593 & 2.4143 & 0.6521 & 0.3430 & -0.6437 & 2.7109 \\ \hline
			k = 54 & -2.4610 & 0.6106 & -0.2736 & -0.1951 & 2.2355 & 2.2063 & 0.5613 & 0.1979 & -0.5379 & 2.4758 \\ \hline
			k = 55 & -2.2205 & 0.5122 & -0.1612 & -0.2964 & 1.9278 & 1.9177 & 0.4849 & 0.1047 & -0.4407 & 2.2380 \\ \hline
			k = 56 & -2.0467 & 0.4190 & -0.0864 & -0.3832 & 1.7195 & 1.7248 & 0.4147 & 0.0499 & -0.3500 & 2.0652 \\ \hline
			k = 57 & -1.8832 & 0.3269 & -0.0408 & -0.4458 & 1.5228 & 1.5431 & 0.3466 & 0.0209 & -0.2624 & 1.9023 \\ \hline
			k = 58 & -1.7422 & 0.2361 & -0.0163 & -0.4647 & 1.3598 & 1.3939 & 0.2776 & 0.0075 & -0.1783 & 1.7613 \\ \hline
			k = 59 & -1.6109 & 0.1506 & -0.0052 & -0.4250 & 1.2107 & 1.2580 & 0.2065 & 0.0021 & -0.1022 & 1.6297 \\ \hline
			k = 60 & -1.4901 & 0.0788 & -0.0012 & -0.3290 & 1.0798 & 1.1398 & 0.1363 & 0.0005 & -0.0425 & 1.5082 \\ \hline
			k = 61 & -1.3775 & 0.0296 & -0.0002 & -0.2031 & 0.9635 & 1.0357 & 0.0743 & 0.0001 & -0.0070 & 1.3949 \\ \hline
			k = 62 & -1.2680 & 0.0057 & -0.0000 & -0.0884 & 0.8577 & 0.9410 & 0.0293 & 0.0000 & 0.0044 & 1.2847 \\ \hline
			k = 63 & -1.2046 & -0.0001 & -0.0000 & -0.0190 & 0.7887 & 0.8859 & 0.0058 & 0.0000 & 0.0024 & 1.2208 \\ \hline
			k = 64 & -0.8286 & -0.0000 & -0.0000 & -0.0006 & 0.5420 & 0.6347 & 0.0002 & -0.0000 & 0.0001 & 0.8423 \\ \hline
			\\ 
		\end{tabular*}
	}
\end{table}

\begin{table}[ht]
	\centering
	\begin{tabular*}{18cm}{@{\extracolsep{\fill}}lcccccccccc}
		\\
		\hline\hline
		\textbf{VEEF} & $n=1(z)$ & $n=2(z)$ & $n=3(z)$ & $n=4(z)$ & $n=5(z)$ & $n=6(z)$ & $n=7(z)$ & $n=8(z)$ & $n=9(z)$ & $n=10(z)$ \\ \hline
		k = 1 & 0.1775 & 12.0178 & -6.9975 & -14.9664 & 9.9504 & 18.6867 & 26.7975 & -6.4060 & -7.6566 & -12.2588 \\ \hline
		k = 2 & -2.2643 & -1.5234 & -4.3044 & -3.0225 & -0.2803 & 0.3081 & 10.0059 & -1.0675 & 2.2818 & 1.8765 \\ \hline
		k = 3 & -2.1919 & -1.9862 & -3.3174 & -2.5496 & -0.7195 & -1.3410 & 2.8208 & 1.3101 & 2.2610 & 1.8036 \\ \hline
		k = 4 & -2.2914 & -1.8776 & -3.2149 & -2.5765 & -0.6959 & -1.3426 & 1.2199 & 2.0494 & 2.2319 & 1.8015 \\ \hline
		k = 5 & -2.2322 & -1.8785 & -3.3700 & -2.5701 & -0.6974 & -1.3663 & 0.8102 & 2.2168 & 2.2603 & 1.8150 \\ \hline
		k = 6 & -2.2283 & -1.8358 & -3.5820 & -2.5918 & -0.6916 & -1.3721 & 0.5920 & 2.2125 & 2.2548 & 1.8156 \\ \hline
		k = 7 & -2.2283 & -1.8039 & -3.7536 & -2.6160 & -0.6876 & -1.3785 & 0.4169 & 2.2036 & 2.2517 & 1.8205 \\ \hline
		k = 8 & -2.2329 & -1.7743 & -3.8570 & -2.6649 & -0.6867 & -1.3830 & 0.2778 & 2.2007 & 2.2421 & 1.8250 \\ \hline
		k = 9 & -2.2311 & -1.7211 & -3.8380 & -2.7352 & -0.6881 & -1.3807 & 0.1838 & 2.1946 & 2.2116 & 1.8141 \\ \hline
		k = 10 & -2.1800 & -1.5995 & -3.6209 & -2.8052 & -0.6879 & -1.3623 & 0.1470 & 2.1582 & 2.1234 & 1.7453 \\ \hline
		k = 11 & -1.9905 & -1.3563 & -3.1178 & -2.7955 & -0.6761 & -1.3179 & 0.1708 & 2.0439 & 1.9173 & 1.5471 \\ \hline
		k = 12 & -1.6039 & -0.9967 & -2.3174 & -2.5827 & -0.6424 & -1.2469 & 0.2382 & 1.8114 & 1.5700 & 1.1917 \\ \hline
		k = 13 & -1.0818 & -0.6009 & -1.3704 & -2.0920 & -0.5866 & -1.1588 & 0.3228 & 1.4336 & 1.1243 & 0.7486 \\ \hline
		k = 14 & -0.5812 & -0.2736 & -0.5117 & -1.4002 & -0.5241 & -1.0737 & 0.3928 & 0.9545 & 0.6872 & 0.3501 \\ \hline
		k = 15 & -0.2097 & -0.0623 & 0.0634 & -0.7092 & -0.4765 & -1.0014 & 0.4287 & 0.4616 & 0.3422 & 0.0708 \\ \hline
		k = 16 & 0.0173 & 0.0474 & 0.2875 & -0.1913 & -0.4540 & -0.9390 & 0.4322 & 0.0208 & 0.1080 & -0.0854 \\ \hline
		k = 17 & 0.1277 & 0.0930 & 0.2211 & 0.0943 & -0.4513 & -0.8798 & 0.4136 & -0.2957 & -0.0265 & -0.1505 \\ \hline
		k = 18 & 0.1601 & 0.1073 & 0.0124 & 0.1814 & -0.4562 & -0.8157 & 0.3839 & -0.4871 & -0.0945 & -0.1535 \\ \hline
		k = 19 & 0.1436 & 0.1097 & -0.1944 & 0.1233 & -0.4539 & -0.7220 & 0.3466 & -0.5429 & -0.1211 & -0.1222 \\ \hline
		k = 20 & 0.0946 & 0.1084 & -0.3066 & -0.0479 & -0.4090 & -0.5109 & 0.2791 & -0.4875 & -0.1227 & -0.0675 \\ \hline
		k = 21 & 0.0380 & 0.1068 & -0.2966 & -0.2540 & -0.2742 & -0.1861 & 0.1731 & -0.3327 & -0.1147 & -0.0112 \\ \hline
		k = 22 & -0.0019 & 0.1061 & -0.1938 & -0.4122 & -0.0700 & 0.1221 & 0.0671 & -0.0733 & -0.1093 & 0.0262 \\ \hline
		k = 23 & 0.0002 & 0.1022 & -0.0660 & -0.4426 & 0.0018 & 0.2056 & 0.0608 & 0.1995 & -0.1064 & 0.0214 \\ \hline
		k = 24 & 0.0252 & 0.0846 & 0.0213 & -0.3330 & -0.0631 & 0.0629 & 0.1258 & 0.3241 & -0.0886 & -0.0112 \\ \hline
		k = 25 & 0.0232 & 0.0520 & 0.0391 & -0.1649 & -0.0567 & -0.0325 & 0.1280 & 0.1973 & -0.0507 & -0.0189 \\ \hline
		k = 26 & 0.0077 & 0.0213 & 0.0004 & -0.0564 & -0.0156 & -0.0189 & 0.0718 & -0.0447 & -0.0171 & -0.0070 \\ \hline
		k = 27 & 0.0021 & 0.0072 & -0.0614 & -0.0163 & -0.0004 & 0.0003 & 0.0254 & -0.1809 & -0.0030 & -0.0018 \\ \hline
		k = 28 & 0.0019 & 0.0028 & -0.1161 & -0.0016 & 0.0016 & 0.0042 & 0.0039 & -0.1689 & -0.0004 & -0.0010 \\ \hline
		k = 29 & 0.0025 & 0.0024 & -0.1449 & 0.0100 & 0.0027 & 0.0033 & -0.0023 & -0.1072 & -0.0007 & -0.0009 \\ \hline
		k = 30 & 0.0017 & 0.0034 & -0.1414 & 0.0226 & 0.0040 & 0.0019 & -0.0013 & -0.0581 & 0.0001 & -0.0007 \\ \hline
		k = 31 & -0.0006 & 0.0053 & -0.1120 & 0.0326 & 0.0048 & 0.0010 & 0.0017 & -0.0135 & 0.0023 & -0.0004 \\ \hline
		k = 32 & -0.0032 & 0.0073 & -0.0750 & 0.0373 & 0.0048 & 0.0006 & 0.0036 & 0.0372 & 0.0044 & -0.0004 \\ \hline
		k = 33 & -0.0039 & 0.0081 & -0.0517 & 0.0394 & 0.0046 & 0.0005 & 0.0036 & 0.0747 & 0.0043 & -0.0003 \\ \hline
		k = 34 & -0.0022 & 0.0069 & -0.0498 & 0.0412 & 0.0041 & 0.0004 & 0.0031 & 0.0831 & 0.0022 & -0.0003 \\ \hline
		k = 35 & 0.0011 & 0.0031 & -0.0548 & 0.0407 & 0.0028 & 0.0005 & 0.0036 & 0.0681 & -0.0001 & -0.0003 \\ \hline
		k = 36 & 0.0047 & -0.0037 & -0.0487 & 0.0286 & 0.0006 & 0.0006 & 0.0047 & 0.0445 & -0.0015 & -0.0003 \\ \hline
		k = 37 & 0.0066 & -0.0104 & -0.0333 & -0.0013 & -0.0019 & 0.0005 & 0.0054 & 0.0219 & -0.0012 & -0.0003 \\ \hline
		k = 38 & 0.0061 & -0.0127 & -0.0203 & -0.0348 & -0.0034 & -0.0001 & 0.0054 & 0.0033 & -0.0001 & -0.0003 \\ \hline
		k = 39 & 0.0037 & -0.0100 & -0.0133 & -0.0520 & -0.0039 & -0.0007 & 0.0050 & -0.0111 & 0.0011 & -0.0001 \\ \hline
		k = 40 & 0.0007 & -0.0052 & -0.0102 & -0.0511 & -0.0038 & -0.0012 & 0.0043 & -0.0214 & 0.0020 & 0.0001 \\ \hline
		k = 41 & -0.0018 & -0.0010 & -0.0092 & -0.0405 & -0.0035 & -0.0013 & 0.0035 & -0.0279 & 0.0023 & 0.0003 \\ \hline
		k = 42 & -0.0034 & 0.0019 & -0.0090 & -0.0270 & -0.0032 & -0.0012 & 0.0025 & -0.0310 & 0.0022 & 0.0005 \\ \hline
		k = 43 & -0.0042 & 0.0036 & -0.0091 & -0.0145 & -0.0028 & -0.0008 & 0.0013 & -0.0311 & 0.0017 & 0.0006 \\ \hline
		k = 44 & -0.0044 & 0.0042 & -0.0089 & -0.0043 & -0.0023 & -0.0002 & 0.0001 & -0.0288 & 0.0010 & 0.0006 \\ \hline
		k = 45 & -0.0040 & 0.0041 & -0.0082 & 0.0026 & -0.0016 & 0.0007 & -0.0009 & -0.0246 & 0.0003 & 0.0006 \\ \hline
		k = 46 & -0.0032 & 0.0037 & -0.0068 & 0.0063 & -0.0009 & 0.0015 & -0.0014 & -0.0193 & -0.0004 & 0.0006 \\ \hline
		k = 47 & -0.0020 & 0.0032 & -0.0050 & 0.0081 & -0.0006 & 0.0019 & -0.0018 & -0.0136 & -0.0010 & 0.0004 \\ \hline
		k = 48 & -0.0006 & 0.0028 & -0.0032 & 0.0097 & -0.0007 & 0.0020 & -0.0024 & -0.0085 & -0.0013 & 0.0003 \\ \hline
		k = 49 & 0.0006 & 0.0024 & -0.0018 & 0.0104 & 0.0001 & 0.0013 & -0.0028 & -0.0047 & -0.0015 & 0.0001 \\ \hline
		k = 50 & 0.0010 & 0.0022 & -0.0009 & 0.0106 & 0.0008 & 0.0006 & -0.0027 & -0.0022 & -0.0014 & -0.0001 \\ \hline
		\\ 
	\end{tabular*}
\end{table}

\begin{table}[tbp]
	\centering
	\scalebox{0.99}{
		\begin{tabular*}{18cm}{@{\extracolsep{\fill}}lcccccccccccc}
			\\
			\hline\hline
			\textbf{VEEF} & $n=1(z)$ & $n=2(z)$ & $n=3(z)$ & $n=4(z)$ & $n=5(z)$ & $n=6(z)$ & $n=7(z)$ & $n=8(z)$ & $n=9(z)$ & $n=10(z)$ \\ \hline
			k = 51 & 0.0009 & 0.0020 & -0.0004 & 0.0131 & 0.0006 & 0.0005 & -0.0021 & -0.0010 & -0.0012 & -0.0001 \\ \hline
			k = 52 & 0.0007 & 0.0018 & -0.0002 & 0.0159 & 0.0002 & 0.0005 & -0.0015 & -0.0004 & -0.0009 & -0.0000 \\ \hline
			k = 53 & 0.0005 & 0.0015 & -0.0001 & 0.0165 & -0.0001 & 0.0004 & -0.0010 & -0.0001 & -0.0007 & 0.0000 \\ \hline
			k = 54 & 0.0003 & 0.0012 & -0.0000 & 0.0149 & -0.0003 & 0.0002 & -0.0007 & -0.0000 & -0.0004 & 0.0000 \\ \hline
			k = 55 & 0.0002 & 0.0008 & -0.0000 & 0.0122 & -0.0005 & 0.0001 & -0.0004 & -0.0000 & -0.0002 & 0.0001 \\ \hline
			k = 56 & 0.0000 & 0.0005 & -0.0000 & 0.0089 & -0.0006 & -0.0000 & -0.0002 & -0.0000 & -0.0001 & 0.0001 \\ \hline
			k = 57 & -0.0001 & 0.0003 & -0.0000 & 0.0057 & -0.0006 & -0.0001 & -0.0001 & -0.0000 & -0.0001 & 0.0001 \\ \hline
			k = 58 & -0.0001 & 0.0001 & -0.0000 & 0.0031 & -0.0006 & -0.0001 & -0.0000 & -0.0000 & -0.0000 & 0.0001 \\ \hline
			k = 59 & -0.0001 & 0.0000 & -0.0000 & 0.0014 & -0.0005 & -0.0001 & -0.0000 & -0.0000 & -0.0000 & 0.0001 \\ \hline
			k = 60 & -0.0001 & 0.0000 & 0.0000 & 0.0004 & -0.0005 & -0.0001 & -0.0000 & -0.0000 & -0.0000 & 0.0001 \\ \hline
			k = 61 & -0.0001 & 0.0000 & -0.0000 & 0.0001 & -0.0004 & -0.0000 & -0.0000 & 0.0000 & -0.0000 & 0.0001 \\ \hline
			k = 62 & -0.0001 & 0.0000 & -0.0000 & 0.0000 & -0.0002 & 0.0000 & -0.0000 & 0.0000 & 0.0000 & 0.0001 \\ \hline
			k = 63 & -0.0000 & -0.0000 & -0.0000 & 0.0000 & -0.0001 & 0.0001 & -0.0000 & 0.0000 & 0.0000 & 0.0001 \\ \hline
			k = 64 & -0.0000 & 0.0000 & -0.0000 & 0.0000 & -0.0000 & 0.0000 & 0.0000 & 0.0000 & 0.0000 & 0.0000 \\ \hline
			\\ 
		\end{tabular*}
	}
\end{table}